\documentstyle[12pt]{article}
\begin{document}

\setlength{\leftmargin}{0.5cm}
\renewcommand{\labelenumi}{\alph{enumi})}
\setcounter{secnumdepth}{3}

%
%
%
%
%
\newcommand{\lpv}{\lambda\phi^4}
\newcommand{\lvpv}{\lambda\varphi^4}

%
%
\newcommand{\Oe}{O(\epsilon)}
\newcommand{\Oeq}{O(\epsilon^2)}
\newcommand{\Oed}{O(\epsilon^3)}
\newcommand{\OedLq}{O\left(\frac{1}{\Lambda^2}\right)}

%
%
\newcommand{\cH}{{\cal H}}
\newcommand{\cHC}{{\cal H}_{\sssty (C)}}
\newcommand{\cL}{{\cal L}}
\newcommand{\cD}{{\cal D}}
\newcommand{\Eo}{E_{\sssty 0}}
\newcommand{\Ee}{E_{\sssty 1}}
\newcommand{\cV}{{\cal V}}             
\newcommand{\VG}{V_{\sssty G}}
\newcommand{\cVG}{{\cal V}_{\sssty G} }
\newcommand{\cVGPx}{{\cal V}_{\sssty G}(\Phi_{\sssty 0},x)}
\newcommand{\cVGP}{{\cal V}_{\sssty G}(\Phi_{\sssty 0})}
\newcommand{\VGpO}{V_{\sssty G}(\phi_{\sssty 0},\Omega)}
\newcommand{\VGnOn}{V_{\sssty G}(0,\Omega_{\sssty 0})}
\newcommand{\VGnn}{V_{\sssty G}(0,0)}
\newcommand{\Vl}{V_{1loop}}
\newcommand{\UG}{U_{\sssty G}}
\newcommand{\UGpO}{U_{\sssty G}(\phi_{\sssty 0},\Omega)}
\newcommand{\cU}{{\cal U}}
\newcommand{\cUG}{{\cal U}_{\sssty G}}
\newcommand{\cUGP}{{\cal U}_{\sssty G}(\Phi_{\sssty 0})}
\newcommand{\cUGPx}{{\cal U}_{\sssty G}(\Phi_{\sssty 0},x)}
\newcommand{\Ul}{U_{\sssty 1loop}}
\newcommand{\dqVGdOq}{\frac{\partial^2V_G}{\partial\Omega^2}}
\newcommand{\dqVGdOdp}{\frac{\partial^2V_G}{\partial\Omega\partial\phi_0}}
\newcommand{\dqVGdpdO}{\frac{\partial^2V_G}{\partial\phi_0\partial\Omega}}
\newcommand{\dqVGdpq}{\frac{\partial^2V_G}{\partial\phi_{\sssty 0}^2}}
\newcommand{\dqVGdoq}{\frac{\partial^2V_G}{\partial\omega^2}}
\newcommand{\dqVGdodp}{\frac{\partial^2V_G}{\partial\omega\partial\phi_0}}
\newcommand{\dqVGdpdo}{\frac{\partial^2V_G}{\partial\phi_0\partial\omega}}
\newcommand{\dqVGdOdo}{\frac{\partial^2V_G}{\partial\Omega\partial\omega}}
\newcommand{\dqVGdodO}{\frac{\partial^2V_G}{\partial\omega\partial\Omega}}
\newcommand{\Gst}{\Gamma_{stat}}
\newcommand{\Ga}{\Gamma}

%
%
\newcommand{\Gaeh}{\Gamma\left(\frac{1}{2}\right)}
\newcommand{\Gameh}{\Gamma\left(-\frac{1}{2}\right)}
\newcommand{\Gadrh}{\Gamma\left(\frac{3}{2}\right)}
\newcommand{\Gadmdh}{\Gamma\left(\frac{3-d}{2}\right)}
\newcommand{\Gaemdh}{\Gamma\left(\frac{1-d}{2}\right)}
\newcommand{\Gadh}{\Gamma\left(\frac{d}{2}\right)}
\newcommand{\ga}{\gamma}

%
%
\newcommand{\IN}{I_{\sssty N}}
\newcommand{\INO}{I_{\sssty N}(\Omega^2)}
\newcommand{\Ie}{I_{\sssty 1}}
\newcommand{\Ien}{I_{\sssty 1}(0)}
\newcommand{\IeO}{I_{\sssty 1}(\Omega^2)}
\newcommand{\Ieo}{I_{\sssty 1}(\omega^2)}
\newcommand{\IeD}{I_{\sssty 1}(\Delta^2)}
\newcommand{\Iee}{I_{\sssty 1}^{\sssty (1)}}
\newcommand{\Iez}{I_{\sssty 1}^{\sssty (2)}}
\newcommand{\In}{I_{\sssty 0}}
\newcommand{\Inn}{I_{\sssty 0}(0)}
\newcommand{\InO}{I_{\sssty 0}(\Omega^2)}
\newcommand{\InOn}{I_{\sssty 0}(\Omega_{\sssty 0}^2)}
\newcommand{\InOe}{I_{\sssty 0}(\Omega_{\sssty 1}^2)}
\newcommand{\InOz}{I_{\sssty 0}(\Omega_{\sssty 2}^2)}
\newcommand{\InOv}{I_{\sssty 0}(\Omega_v^2)}
\newcommand{\Ino}{I_{\sssty 0}(\omega^2)}
\newcommand{\InD}{I_{\sssty 0}(\Delta^2)}
\newcommand{\Inmr}{I_{\sssty 0}(m_{\sssty R}^2)}
\newcommand{\Inm}{I_{\sssty 0}(\mu^2)}
\newcommand{\Ine}{I_{\sssty 0}^{\sssty (1)}}
\newcommand{\Inz}{I_{\sssty 0}^{\sssty (2)}}
\newcommand{\Ineq}{I_{\sssty 0}^{{\sssty (1)}2}}
\newcommand{\Inzq}{I_{\sssty 0}^{{\sssty (2)}2}}
\newcommand{\Inxn}{I_{\sssty 0}(x_{\sssty 0})}
\newcommand{\Inxv}{I_{\sssty 0}(x_v)}
\newcommand{\Inyv}{I_{\sssty 0}(y_v)}
\newcommand{\Imi}{I_{\sssty -1}}
\newcommand{\ImiO}{I_{\sssty -1}(\Omega^2)}
\newcommand{\ImiOn}{I_{\sssty -1}(\Omega_{\sssty 0}^2)}
\newcommand{\ImiOv}{I_{\sssty -1}(\Omega_v^2)}
\newcommand{\Imio}{I_{\sssty -1}(\omega^2)}
\newcommand{\ImiD}{I_{\sssty -1}(\Delta^2)}
\newcommand{\Imimr}{I_{\sssty -1}(m_{\sssty R}^2)}
\newcommand{\Imim}{I_{\sssty -1}(\mu^2)}
\newcommand{\Imix}{I_{\sssty -1}(x)}
\newcommand{\Imixn}{I_{\sssty -1}(x_{\sssty 0})}
\newcommand{\Imixv}{I_{\sssty -1}(x_v)}
\newcommand{\Imiyv}{I_{\sssty -1}(y_v)}
\newcommand{\eseI}{(1+6\eta)I_{\sssty -1}}
\newcommand{\eseIm}{(1+6\eta)I_{\sssty -1}(\mu^2)}
\newcommand{\aedI}{\frac{8\eta}{I_{\sssty -1}}}
\newcommand{\Idae}{\frac{I_{\sssty -1}}{8\eta}}
\newcommand{\eaeI}{(1+8\eta)I_{\sssty -1}}
\newcommand{\eae}{(1+8\eta)}
\newcommand{\eveI}{(1+4\eta)I_{\sssty -1}}
\newcommand{\eve}{(1+4\eta)}
\newcommand{\dsze}{(3+16\eta)}

%
%
\newcommand{\dmdrh}{\frac{d-3}{2}}
\newcommand{\drmdh}{\frac{3-d}{2}}
\newcommand{\dmeh}{\frac{d-1}{2}}
\newcommand{\emdh}{\frac{1-d}{2}}
\newcommand{\deh}{\frac{d}{2}}

%
%
\newcommand{\fh}{\hat{f}}
\newcommand{\fhx}{\hat{f}(x)}
\newcommand{\fhxn}{\hat{f}(x_{\sssty 0})}
\newcommand{\fhy}{\hat{f}(y)}
\newcommand{\fhxe}{\hat{f}'(x)}
\newcommand{\fhxve}{\hat{f}'(x_{v})}
\newcommand{\fhxne}{\hat{f}'(x_{\sssty 0})}
\newcommand{\fhye}{\hat{f}'(y)}
\newcommand{\fhyve}{\hat{f}'(y_{v})}
\newcommand{\fhxz}{\hat{f}''(x)}
\newcommand{\fhxvz}{\hat{f}''(x_{v})}
\newcommand{\fhxnz}{\hat{f}''(x_{\sssty 0})}
\newcommand{\fhyz}{\hat{f}''(y)}
\newcommand{\fhyvz}{\hat{f}''(y_{v})}
%

%
%
\newcommand{\xv}{\vec{x}}
\newcommand{\yv}{\vec{y}}
\newcommand{\zv}{\vec{z}}
\newcommand{\kv}{\vec{k}}
\newcommand{\kvb}{|\vec{k}|}
\newcommand{\pv}{\vec{p}\,}
\newcommand{\kvq}{{\vec{k}}^2}
\newcommand{\kvSq}{{\vec{k}}'^2}
\newcommand{\pvq}{{\vec{p}}^{\,2}}
\newcommand{\kne}{k_{\sssty 0}^{\sssty (1)}}
\newcommand{\kneS}{k_{\sssty 0}^{'\sssty (1)}}
\newcommand{\kneq}{k_{\sssty 0}^{{\sssty (1)}2}}
\newcommand{\knz}{k_{\sssty 0}^{\sssty (2)}}
\newcommand{\knzS}{k_{\sssty 0}^{'\sssty (2)}}
\newcommand{\knzq}{k_{\sssty 0}^{{\sssty (2)}2}}
%

%
%
\newcommand{\La}{\Lambda}
\newcommand{\Laq}{\Lambda^2}

%
%
\newcommand{\muq}{\mu^2}
\newcommand{\mqdOq}{\frac{\muq}{\Omega^2}}
\newcommand{\mqdOqheh}{\left(\frac{\muq}{\Omega^2}\right)^{\frac{\epsilon}{2}}}
\newcommand{\mqdvpheh}{\left(\frac{\muq}{4\pi}\right)^{\frac{\epsilon}{2}}}
\newcommand{\Om}{\Omega}
\newcommand{\Omh}{\frac{\Omega}{2}}     
\newcommand{\Omn}{\Omega_{\sssty 0}}
\newcommand{\Omnq}{\Omega_{\sssty 0}^{2}}
\newcommand{\Omvq}{\Omega_{v}^{2}}
\newcommand{\Omq}{\Omega^2}
\newcommand{\Omqh}{\frac{\Omega^2}{2}}
\newcommand{\Omv}{\Omega^4}
\newcommand{\Ome}{\Omega_{\sssty 1}}
\newcommand{\Omeq}{\Omega_{\sssty 1}^2}
\newcommand{\Omz}{\Omega_{\sssty 2}}
\newcommand{\Omzq}{\Omega_{\sssty 2}^2}
\newcommand{\lnOqmq}{\ln\frac{\Omq}{\mu^2}}
\newcommand{\lnOqmrq}{\ln\frac{\Omq}{m_{\sssty R}^2}}
\newcommand{\lnmrqmq}{\ln\frac{m_{\sssty R}^2}{\mu^2}}
\newcommand{\Dl}{\Delta}
\newcommand{\Dlq}{\Delta^2}
\newcommand{\om}{\omega}
\newcommand{\omn}{\omega_{\sssty 0}}
\newcommand{\omnq}{\omega_{\sssty 0}^{2}}
\newcommand{\omvq}{\omega_{v}^{2}}
\newcommand{\omq}{\omega^2}
\newcommand{\omqh}{\frac{\omega^2}{2}}
\newcommand{\omv}{\omega^4}
\newcommand{\omk}{\omega_{\sssty\kv}}
\newcommand{\omp}{\omega_{\sssty\pv}}
\newcommand{\xn}{x_{\sssty 0}}
\newcommand{\xdel}{x_{\delta}}   
\newcommand{\wx}{\sqrt{x}}
\newcommand{\wepx}{\sqrt{1+x}}
\newcommand{\wepxd}{\sqrt{(1+x)^3}}
\newcommand{\wepxn}{\sqrt{1+x_{\sssty 0}}}
\newcommand{\wepxv}{\sqrt{1+x_v}}
\newcommand{\lnfxxn}{\ln\left(
               \frac{\sqrt{x_{\sssty 0}}(1+\sqrt{1+x})}
                    {\sqrt{x}(1+\sqrt{1+x_{\sssty 0}})}\right)}
\newcommand{\lnfx}{\ln\left(\frac{1+\sqrt{1+x}}{\sqrt{x}}\right)}
\newcommand{\lnfxn}{\ln\left(\frac{1+\sqrt{1+x_{\sssty 0}}}
                                  {\sqrt{x_{\sssty 0}}}\right)}
\newcommand{\lnfxv}{\ln\left(\frac{1+\sqrt{1+x_v}}{\sqrt{x_v}}\right)}
\newcommand{\xh}{\frac{x}{2}}
\newcommand{\dxdPhioq}{\frac{dx}{d\Phi_{\sssty 0}^2}}
\newcommand{\dydPhioq}{\frac{dy}{d\Phi_{\sssty 0}^2}}

%
%
\newcommand{\zwe}{12\eta}
\newcommand{\emze}{(1-12\eta)}
\newcommand{\etmax}{\eta_{max}}
\newcommand{\etmie}{\eta_{min,\sssty 1}}
\newcommand{\etmiz}{\eta_{min,\sssty 2}}
\newcommand{\etmae}{\eta_{max,\sssty 1}}
\newcommand{\etmaz}{\eta_{max,\sssty 2}}

%
%
\newcommand{\eps}{\epsilon}
\newcommand{\epsh}{\frac{\epsilon}{2}}
\newcommand{\eme}{(1-\epsilon)}

%
%
\newcommand{\zFe}{\left._{\sssty 2}F_{\sssty 1}\right.}

%
%
\newcommand{\Jx}{J(\vec{x})}
\newcommand{\Jy}{J(\vec{y})}

%
%
\newcommand{\ka}{\kappa}
\newcommand{\lbd}{\lambda}
\newcommand{\lba}{\lambda_{\sssty B}}
\newcommand{\lre}{\lambda_{\sssty R}}
\newcommand{\lbh}{\hat{\lambda}_{\sssty B}}
\newcommand{\lbhcr}{\hat{\lambda}_{{\sssty B,}crit}}
\newcommand{\eba}{e_{\sssty B}}
\newcommand{\ebq}{e_{\sssty B}^2}
\newcommand{\ebqe}{\frac{e_{\sssty B}^2}{1-\epsilon}}

%
%
\newcommand{\mba}{m_{\sssty B}}
\newcommand{\mbq}{m_{\sssty B}^2}
\newcommand{\mbhq}{\hat{m}_{\sssty B}^2}
\newcommand{\mre}{m_{\sssty R}}
\newcommand{\mrq}{m_{\sssty R}^2}
\newcommand{\mhq}{\hat{m}^2}
\newcommand{\mrhq}{\hat{m}_{\sssty R}^2}
\newcommand{\MH}{M_{\mbox{\tiny Higgs}}}

%
%
\newcommand{\piq}{\pi^2}
\newcommand{\tet}{\theta}
\newcommand{\vph}{\varphi}
\newcommand{\vecph}{\vec{\varphi}}
\newcommand{\phx}{\varphi{(\vec{x})}}
\newcommand{\Phx}{\phi{(\vec{x})}}
\newcommand{\Phy}{\phi{(\vec{y})}}
\newcommand{\PhSx}{\phi'{(\vec{x})}}
\newcommand{\PhSy}{\phi'{(\vec{y})}}
\newcommand{\Phiet}{\Phi_{\sssty 1}(\theta)}
\newcommand{\Phizt}{\Phi_{\sssty 2}(\theta)}
\newcommand{\pie}{\pi_{\sssty 1}}
\newcommand{\piz}{\pi_{\sssty 2}}
\newcommand{\pht}{\tilde{\phi}}
\newcommand{\phie}{\varphi_{\sssty 1}}
\newcommand{\phiz}{\varphi_{\sssty 2}}
\newcommand{\vpht}{\tilde{\varphi}}
\newcommand{\vphte}{\tilde{\varphi}_{\sssty 1}}
\newcommand{\vphtz}{\tilde{\varphi}_{\sssty 2}}
\newcommand{\vphtep}{\dot{\tilde{\varphi}}_{\sssty 1}}
\newcommand{\vphtzp}{\dot{\tilde{\varphi}}_{\sssty 2}}
\newcommand{\phio}{\phi_{\sssty 0}}
\newcommand{\phioe}{\phi_{\sssty 0,1}}
\newcommand{\phioz}{\phi_{\sssty 0,2}}
\newcommand{\phioq}{\phi_{\sssty 0}^{2}}
\newcommand{\phiomin}{\phi_{{\sssty 0},min}}
\newcommand{\phioqcr}{\phi_{{\sssty 0},crit}^2}
\newcommand{\phioqcre}{\phi_{{\sssty 0},crit,\sssty 1}^2}
\newcommand{\phioqcrz}{\phi_{{\sssty 0},crit,\sssty 2}^2}
\newcommand{\phiox}{\phi_{\sssty 0}{(\vec{x})}}
\newcommand{\phioy}{\phi_{\sssty 0}{(\vec{y})}}
\newcommand{\phior}{\phi_{\sssty 0,R}}
\newcommand{\phiorq}{\phi_{\sssty 0,R}^{2}}
\newcommand{\phirq}{\phi_{\sssty R}^{2}}
\newcommand{\Phio}{\Phi_{\sssty 0}}
\newcommand{\Phiocr}{\Phi_{{\sssty 0},crit}}
\newcommand{\Phiocre}{\Phi_{{\sssty 0},crit,\sssty 1}}
\newcommand{\Phiomin}{\Phi_{{\sssty 0},min}}
\newcommand{\Phiomax}{\Phi_{{\sssty 0},max}}
\newcommand{\Phioq}{\Phi_{\sssty 0}^{2}}
\newcommand{\PhioSq}{\Phi_{\sssty 0}^{'2}}
\newcommand{\Phioqcr}{\Phi_{{\sssty 0},crit}^{2}}
\newcommand{\Phioqcre}{\Phi_{{\sssty 0},crit,\sssty 1}^2}
\newcommand{\Phioqcrz}{\Phi_{{\sssty 0},crit,\sssty 2}^2}
\newcommand{\vhq}{\hat{v}^2}
\newcommand{\vh}{\hat{v}}
\newcommand{\vhqcl}{\hat{v}^2_{cl}}
\newcommand{\vqcl}{v^2_{cl}}
\newcommand{\psio}{\psi_{\sssty 0}}
\newcommand{\Psio}{\Psi_{\sssty 0}}
\newcommand{\psiG}{\psi_{\sssty G}}
\newcommand{\PsiG}{\Psi_{\sssty G}}
\newcommand{\psiC}{\psi_{\sssty (C)}}
\newcommand{\psibra}{\langle\,\psi\,}
\newcommand{\psiket}{|\,\psi\,\rangle}
\newcommand{\vacbra}{\langle\,0\,}   
\newcommand{\vacket}{|\,0\,\rangle}  
\newcommand{\si}{\sigma}
\newcommand{\six}{\sigma(x)}
\newcommand{\th}{\vartheta}
\newcommand{\thx}{\vartheta(x)}
\newcommand{\xtd}{\tilde{x}}    
\newcommand{\ytd}{\tilde{y}}    

%
%
\newcommand{\Dxy}{D(\xv-\yv\,)}
\newcommand{\gxy}{g(\xv,\yv\,)}
\newcommand{\hxy}{h(\xv,\yv\,)}
\newcommand{\ft}{\tilde{f}}
\newcommand{\ftp}{\tilde{f}(\pv)}
\newcommand{\ftk}{\tilde{f}(\kv)}
\newcommand{\ad}{a^{\dagger}}
\newcommand{\bd}{b^{\dagger}}                
\newcommand{\vd}{v^{\dagger}}                
\newcommand{\ud}{u^{\dagger}}                
\newcommand{\dg}{\dagger}                    
\newcommand{\hepsk}{\hat{\epsilon}_{\kv}}

%
%
\newcommand{\Amu}{A_{\sssty\mu}}
\newcommand{\Amo}{A^{\sssty\mu}}
\newcommand{\Anu}{A_{\sssty\nu}}
\newcommand{\Ano}{A^{\sssty\nu}}
\newcommand{\Fmno}{F^{\sssty\mu\nu}}
\newcommand{\Fmnu}{F_{\sssty\mu\nu}}
\newcommand{\Piv}{\vec{\Pi}}
\newcommand{\Pivt}{\vec{\Pi}^{t}}
\newcommand{\Pivtq}{\vec{\Pi}^{t\,2}}
\newcommand{\Pivl}{\vec{\Pi}^{\ell}}
\newcommand{\Bv}{\vec{B}}
\newcommand{\Av}{\vec{A}}
\newcommand{\Avt}{\vec{A}^{t}}
\newcommand{\Avtq}{\vec{A}^{t\,2}}
\newcommand{\jv}{\vec{j}}
\newcommand{\etav}{\vec{\eta}}
\newcommand{\etavq}{\vec{\eta}^{\,2}}

%
%
\newcommand{\ra}{\,\rangle}
\newcommand{\la}{\langle\,}
\newcommand{\ptd}{\partial}
\newcommand{\ptdmu}{\partial_{\sssty\mu}}
\newcommand{\ptdmo}{\partial^{\sssty\mu}}
\newcommand{\ptdnu}{\partial_{\sssty\nu}}
\newcommand{\dqdpq}{\frac{\delta^2}{\delta\phi^2}}
\newcommand{\dqdpxq}{\frac{\delta^2}{\delta\phi(\vec{x})^2}}
\newcommand{\dqdpSxq}{\frac{\delta^2}{\delta\phi'(\vec{x})^2}}
\newcommand{\nab}{\vec{\nabla}}
\newcommand{\mnqpmq}{\left(-\vec{\nabla}^2+m^2\right)}

%
%
\newcommand{\Fint}{\int {\cal D}}
\newcommand{\Fintph}{\int {\cal D}\phi\,}
\newcommand{\Intne}{\int_0^1}
\newcommand{\Intnedx}{\int_0^1 dx\,}
\newcommand{\Intnedy}{\int_0^1 dy\,}
\newcommand{\Intddx}{\int d^3\!x\,}
\newcommand{\ddx}{d^3\!x\,}
\newcommand{\Intddy}{\int d^3\!y\,}
\newcommand{\ddy}{d^3\!y\,}
\newcommand{\Intddk}{\int d^3\!k\,}
\newcommand{\Intddkp}{\int\frac{d^3\!k}{(2\pi)^3}\,}
\newcommand{\Intddkzp}{\int\frac{d^3\!k}{2(2\pi)^3}\,}
\newcommand{\Intdnkzp}{\int\frac{d^{\nu}\!k}{2(2\pi)^{\nu}}\,}
\newcommand{\IntdDkzp}{\int\frac{d^{d}\!k}{2(2\pi)^{d}}\,}
\newcommand{\IntnLdk}{\int_0^{\La}\frac{dk}{(2\pi)^2}\,}
\newcommand{\dkOe}{(dk)_{\Omega_1}}
\newcommand{\dkOz}{(dk)_{\Omega_2}}
\newcommand{\del}{\delta}
\newcommand{\deld}{\delta^{\sssty (3)}}

%
%
\newcommand{\eh}{\frac{1}{2}}
\newcommand{\ev}{\frac{1}{4}}
\newcommand{\drh}{\frac{3}{2}}

%
%
\newcommand{\be}{\begin{equation}}
\newcommand{\ee}{\end{equation}}
\newcommand{\ba}{\begin{array}}
\newcommand{\ea}{\end{array}}
\newcommand{\bea}{\begin{eqnarray}}
\newcommand{\eea}{\end{eqnarray}}
\newcommand{\mtext}[1]{\mbox{\rm #1}}
\newcommand{\lk}{\left}
\newcommand{\rk}{\right}
\newsavebox{\TRS}
\sbox{\TRS}{\hspace{.5em} = \hspace{-1.8em}
                 \raisebox{1ex}{\mbox{\scriptsize TRS}} }
\newcommand{\eqtrs}{\usebox{\TRS}}
\newsavebox{\defgleich}
\sbox{\defgleich}{\ :=\ }
\newcommand{\eqdef}{\usebox{\defgleich}}
\newsavebox{\LSIM}
\sbox{\LSIM}{\raisebox{-1ex}{$\ \stackrel{\textstyle<}{\sim}\ $}}
\newcommand{\lsim}{\usebox{\LSIM}}
\newsavebox{\GSIM}
\sbox{\GSIM}{\raisebox{-1ex}{$\ \stackrel{\textstyle>}{\sim}\ $}}
\newcommand{\gsim}{\usebox{\GSIM}}
\newcommand{\lrar}{\longrightarrow}
\newcommand{\rar}{\rightarrow}
\newcommand{\Rar}{\Rightarrow}
\newcommand{\itdot}{\hspace*{2cm}$\bullet$\ \ }

%
%
\newcommand{\tsty}{\textstyle}
\newcommand{\ssty}{\scriptstyle}
\newcommand{\sssty}{\scriptscriptstyle}
\newcommand{\fns}{\footnotesize}

%
%
\newsavebox{\Dot}
\sbox{\Dot}{\raisebox{0.5ex}{\rule{0.400pt}{0.400pt}}}
\newsavebox{\Dotspc}
\sbox{\Dotspc}{\raisebox{0.5ex}{\rule{0.400pt}{0.400pt}}\hspace{2.009pt}}
\newcommand{\lone}{\raisebox{0.5ex}{\rule{7.227pt}{0.400pt}}}
\newcommand{\ltwo}{\usebox{\Dotspc}\usebox{\Dotspc}\usebox{\Dotspc}\usebox{\Dot}}
\newcommand{\lthr}{\raisebox{0.5ex}{\rule{7.227pt}{0.800pt}}}

%
%
\newcommand{\QM}{Quantenmechanik}
\newcommand{\QFT}{Quantenfeldtheorie}
\newcommand{\Gss}{Gau"s}
\newcommand{\Ld}{Lagrangedichte}
\newcommand{\Ho}{Hamiltonoperator}
\newcommand{\Hd}{Hamiltondichte}
\newcommand{\Sch}{Schr"odinger}
\newcommand{\DGL}{Differentialgleichung}
\newcommand{\Gz}{Grundzustand}
\newcommand{\EWW}{Erwartungswert}
\newcommand{\VEW}{Vakuumerwartungswert}
\newcommand{\BA}{bare}
\newcommand{\OPG}{Optimierungsgleichung}
\newcommand{\Sy}{Symmetrie}
\newcommand{\sy}{symmetrisch}

\setlength{\baselineskip}{12pt}
\title{
Effective Hamiltonian for Scalar Theories in the Gaussian Approximation}
\author{H.W.L. Naus,
T. Gasenzer and H.--J. Pirner
\\\\
Institut f\"ur Theoretische Physik\\
Universit\"at Heidelberg, Germany}

\date{\today}

\vspace{1.0cm}

\maketitle

\begin{abstract}
\setlength{\baselineskip}{24pt}
\noindent
{We use a Gaussian wave functional for the ground state to reorder
the Hamiltonian into a free part with a variationally determined
mass and the rest. Once spontaneous symmetry breaking is taken into
account, the residual Hamiltonian can, in principle, be treated perturbatively.
In this scheme we analyze the $O(1)$ and $O(2)$ scalar models.
For the $O(2)$--theory we first explicitly calculate the massless
Goldstone excitation and then show that the one-loop corrections of
the effective Hamiltonian do not generate a mass.\\
}
\end{abstract}
\vfill{\rm\normalsize{HD--TVP--95--11}}
\newpage

\setlength{\baselineskip}{12pt}

\section{Introduction}
Hamiltonian field theory has been the starting point of modern quantum field
theory, but since then has been less used than Lagrangian field theory or
action oriented approaches. The advantage of a Hamiltonian formulation
\cite{Jackiw90} is the appearance of "wave functions", which are functionals of
the fields and allow an intuitive understanding of the ground state. Recently,
for the analysis of gauge theories \cite{Lenz94} and light--cone theory
\cite{Perry90} a Hamiltonian treatment has been revived.  The concept of an
effective Hamiltonian, as
exploited in this paper, has actually been developed  recently in the context
of
one--dimensional light--cone theories \cite{Prok}.
There exists a substantial literature \cite{Stevenson84,Kerman1} on the use of
variational treatments in Hamiltonian field theory. One even finds quantitative
estimates of the Higgs--mass, based on this technique \cite{Consoli92}. The
variational technique is inherently nonperturbative, so one hopes to capture
features which go beyond the standard loop expansion.
In many--body theory the Gaussian wave functional with an effective mass
corresponds to the self--consistent mean field theory. In fact, also more
sophisticated methods of many--body theory like the cluster expansion
\cite{Bishop87} or the RPA--approximation can be used to improve the mean field
result.

In this paper we treat $O(N)$--models for $N=1,2$ to analyse two main features,
($i$) the bounds on the mass $m(1)$ of the massive scalar after
renormalization, and ($ii$) the mass $m(2)$ of the Goldstone particle.
In the recent literature \cite{Consoli92} a Higgs--mass of 2 TeV has been
"predicted" for the experimentally given vacuum expectation value of the Higgs
field in the Standard Model. We use a variational treatment of the scalar
sector with a finite cutoff and present results different from the previous
estimates in Hamiltonian field theory: $\MH\le 1.7$ TeV.

In previous Hamiltonian based work on the $O(2)$--model a finite Goldstone
boson mass $m(2)\simeq m(1)/2.06$ \cite{Stevenson87} is obtained, in
contradiction to Goldstone's theorem. We will show that a symmetric Gaussian
ansatz and a further rediagonalization of the effective Hamiltonian exactly
gives a zero mass Goldstone excitation. This result forms the basis for any
further exploration of the Higgs model or more complicated gauge theories like
QCD.
Possible problems due to the violation of local gauge invariance, which in
general appear in approximative treatments, will be avoided by starting with
"gauge fixed Hamiltonians" from which the redundant degrees of freedom have
been eliminated  \cite{Lenz94}.

There has been an extensive discussion of the Gaussian Effective Potential
(GEP) \cite{Stevenson84}. The emphasis of that work has been to calculate the
energy of the vacuum as a function of the symmetry breaking zero mode of the
field. In general the mass gap or the spectrum of particles, however, is the
more interesting phenomenological quantity. Therefore one has to address the
problem of calculating the energy and the dispersion relation of excited
states. From our point of view the Gaussian wave functional presents an
efficient way of reordering the Hamiltonian into quadratic and higher
polynomial parts. Our approach
is time--independent; recently, also time--dependent variational equations
have been investigated in $\phi^4$ field theory \cite{Kerman2}.

In $0+1$ dimensions, i.e. quantum mechanics, the anharmonic oscillator is an
elucidating example. It shows the problems we are facing in the case of
spontaneous symmetry breaking (SSB) in the $O(2)$--model. The ground state
solution of $H$
\be
   H = \eh\lk(p^2 + m^2x^2\rk) + \lbd x^4,
\ee
is approximated by the Gaussian wave function
\be
   \phi_G(x) = \left(\frac{\Om}{\pi}\right)^{\eh} \exp\lk[-\eh\Om(x-\xn)^2\rk].
\ee
Minimization of the expectation value of the Hamiltonian operator,
\bea
   \VG(\xn,\Om)
   &=& \la\phi_G|H|\phi_G\ra \nonumber\\
   &=& \ev\Om+\frac{m^2}{4\Om}+\eh m^2\xn^2+
       \lbd\lk[\frac{3}{\Om}\xn^2+\frac{3}{4\Omq}+\xn^4\rk]
   \label{eq:Vgau}
\eea
with respect to $\xn$ and $\Om$ gives the equations
\bea
\xn\left(m^2+\frac{6\lambda}{\Omega}+4\lambda\xn^2\right) &=& 0  , \nonumber \\
\Omega^2-m^2-12\lambda\xn^2-\frac{6\lambda}{\Omega} &=& 0 .
   \label{eq:vari}
\eea
For the case $\xn\not=0$ one explicitly finds
\be
   \Omq = 8\lbd\xn^2.
   \label{eq1:Omega}
\ee
Equivalently, one defines the trial ground state via creation and annihilation
operators $\ad_{\Om}$ and $a_{\Om}$,
\bea
a_{\Om}|\phi_G\ra &=& 0  ,\nonumber \\
\la\phi_G|\phi_G\ra  &=& 1, \nonumber \\
\la\phi_G|\,x\,|\phi_G\ra &=& \xn.
\eea
Note that in this way excited states are also implicitly defined.
Coordinate and momentum are correspondingly decomposed
\be
   p=-i\sqrt{\frac{\Om}{2}}\lk(a_{\Om}-\ad_{\Om}\rk)\quad \mtext{and}\quad
   \tilde{x} := x-\xn = \sqrt{\frac{1}{2\Om}}\lk(a_{\Om}+\ad_{\Om}\rk).
\ee
Normal ordering with respect to $a_{\Om}$ and $\ad_{\Om}$, which is denoted
by $ : \; : $, yields
\bea
  :x^2: &=& x^2 - \frac{1}{2\Omega} , \nonumber\\
  :x^3: &=& x^3 - \frac{3}{2\Omega}:x: , \nonumber\\
  :x^4: &=& x^4 - \frac{6}{2\Omega}:x^2: - 3\left(\frac{1}{2\Omega}\right)^2
 , \nonumber\\
  :p^2: &=& p^2 - \frac{\Omega}{2} .
\eea
The normal ordered Hamiltonian reads
\bea
   H &=& :\eh\lk(p^2 + m^2x^2\rk) + \lbd x^4   \nonumber\\
     & & +\ \frac{\Omega}{4} +\frac{m^2}{4\Omega} +\frac{3\lbd}{\Omega}x^2
+ \frac{3\lbd}{4\Omega^2} : .
\eea
Herewith one easily verifies eq. (\ref{eq:Vgau}).
Now we consider the  difference $H_R=H-\VG$ and using eqs. (\ref{eq:vari})
we obtain
\bea
   H_R &=& :\eh\lk(p^2 + (m^2+\frac{6\lambda}{\Omega})(x^2-\xn^2)\rk)
+ \lbd(x^4-\xn^4): \nonumber\\
   &=& :\eh\lk(p^2 + (\Omega^2-12\lbd\xn^2)(x^2-\xn^2)\rk)
+ \lbd(x^4-\xn^4): \nonumber\\
   &=& :\eh\lk(p^2 + \Omega^2\tilde{x}^2\rk)
+ \lbd\tilde{x}^4 + 4\lbd\xn\tilde{x}^3: .
\eea
$H_R$ can be rewritten
in terms of the creation and annihilation operators
\be
 H_R = H_0 + H_I ,
\ee
with
\bea
 H_0 &=& \Om\ad_{\Om} a_{\Om} ,   \\
  H_I &=& 4\lbd\xn\left(\frac{1}{2\Om}\right)^{\frac{3}{2}}
           \lk(a_{\Om}^3+3\ad_{\Om} a_{\Om}^2+3a_{\Om}^{\dg 2}a_{\Om}+
a_{\Om}^{\dg 3}\rk) \nonumber\\
     & & +\ \lbd\left(\frac{1}{2\Om}\right)^{2}
           \lk(a_{\Om}^4+4\ad_{\Om} a_{\Om}^3+6a_{\Om}^{\dg 2}a_{\Om}^2+
4a_{\Om}^{\dg 3}a_{\Om}+a_{\Om}^{\dg 4}\rk),
\label{eq:Hnullqm}
\eea
where we separated the harmonic part $H_0$.
The remaining part of the Hamiltonian, i.e. $H_I$, will be treated in
perturbation theory. We will see that
the use of perturbation theory is well justified if $\Om$ is large.
This is obvious for $\xn=0$, where only the quartic
interaction is present. In the "broken" phase, however, there is
a cubic interaction with an effective coupling constant containing
 $\xn$. An explicit (second order) calculation shows that
for localized trial states the relevant gap between the first excited state
$\Ee$ and the ground state $\Eo$ is dominated by the splitting of energy levels
for the unperturbed Hamiltonian
\be
  \Ee-\Eo = \Om\,\lk(1-\frac{6\lbd}{\Om^3}\rk).
\ee
In our system of units $H$, $m$ and $\Om$ have units of energy,
$[H]=[m]=[\Om]=E$, $[x]=1/\sqrt{E}$ and $[\lbd]=E^3$. One can see the
intuitively obvious fact that the corrections are small for large $\Om$. To be
more specific, $\lbd/\Om^3$ has to be smaller than unity for perturbation
theory to hold. In the opposite case the Gaussian approach fails and higher
clusters play an important role.

For the two--dimensional problem
in quantum mechanics, a similar variational treatment is possible.
The Hamiltonian of this $O(2)$--model reads
\bea
     H &=& \eh\lk(p_x^2+p_y^2\rk)+\eh m^2\lk(x^2+y^2\rk)
           +\lbd\lk(x^2+y^2\rk)^2.
\eea
Here, in quantum mechanics, the best approach is to rewrite the Hamiltonian
in spherical variables and solve for the radial wave function.
However, this cannot be easily generalized to quantum field theory and
therefore we will stick to Cartesian coordinates.
Our main new idea is to use a Gaussian wave function
\be
  \phi_{G}(x,y) = \left(\frac{\Om}{\pi}\right)^{\eh}
           \exp\left[-\eh\Om(x-\xn)^2-\eh\Om y^2\right],
\ee
with the {\it same} size parameter in longitudinal and transverse direction.
This will be seen to be a good starting point for quantum field theory,
because it preserves Goldstone's theorem for higher dimensions of space--time.
Thus the same parameter $\Om$ appears in the decomposition of
coordinates and momenta in two dimensions
\bea
 p_x=-i\sqrt{\Omh}\lk(a_{\Om}-\ad_{\Om}\rk),&\qquad&
\xtd=\sqrt{\frac{1}{2\Om}}\lk(a_{\Om}+\ad_{\Om}\rk), \nonumber\\
 p_y=-i\sqrt{\Omh}\lk(b_{\Om}-\bd_{\Om}\rk),&\qquad&
   y=\sqrt{\frac{1}{2\Om}}\lk(b_{\Om}+\bd_{\Om}\rk).\nonumber
\eea
The same variational treatment as before leads to the normal ordered form
of the effective Hamiltonian, $H_R=H-V_G=H_0+H_I$,
\bea
    H_0 &=& :\eh p_x^2+\eh(2\Omq)\xtd^2+\eh p_y^2: , \nonumber\\
    H_I  &=& :4\lbd\xn\xtd^3+\lbd\xtd^4+2\lbd y^2\xtd^2+4\lbd\xn\xtd y^2
             +\lbd y^4:.
\eea
Only the more interesting case $\xn \ne 0$ is considered.
Because of the appearance of $2\Omq = 8\lbd\xn^2$, with
$\xn:=[(-m^2-8\lbd/\Om)/4\lbd]^{1/2}$, in front of the quadratic term $\xtd^2$
and the vanishing of the term purely quadratic in $y$, the normal ordered form
$H_0$ has not the usual representation in terms of the occupation number
operators $\ad_{\Om} a_{\Om}$
and $\bd_{\Om} b_{\Om}$.
It is therefore necessary to introduce two Bogoliubov transformations for the
pair of operators $a_{\Om}$, $\ad_{\Om}$ and $b_{\Om}$, $\bd_{\Om}$ in order to
diagonalize $H_0$.
To preserve the commutation relations one has
\bea
   u &=& \cosh\alpha\,a_{\Om} + \sinh\alpha\,\ad_{\Om} ,\nonumber\\
   v &=& \cosh\beta\,b_{\Om} + \sinh\beta\,\bd_{\Om} ,
\eea
where $\alpha$ and $\beta$ are determined from commuting $u$ and $v$ with
$H_0$:
\bea
  \lk[u,H_0\rk] &=& \om_1 u, \nonumber\\
  \lk[v,H_0\rk] &=& \om_2 v.
\eea
The eigenvalues of these commutation relations are indeed the physical
energy eigenvalues.
One finds $\tanh\alpha = 3-2\sqrt{2}$, $\tanh\beta=-1$ and
\bea
   \om_1 &=& \Om\sqrt{2} , \nonumber\\
\om_2 &=& 0.
\eea
The zero eigenvalue for the $y$--oscillator is related to the fact that the
effective classical potential is flat after spontaneous symmetry breaking in
$x$--direction.
The Bogoliubov transformation, however, is formally not well defined, since
we have $\beta=-\infty$. This reflects the fact that there is no unitary
transformation between non--normalizable plane waves  and normalizable
oscillator wave functions. In section (4) we will regularize the theory
such that the second frequency is non--zero and, consequently,
the wave function is normalizable. This also renders the Bogoliubov
transformation to be well behaved, i.e. unitary. Furthermore, the
perturbative corrections due to $H_I$ will yield a finite frequency,
even in the limit where the regulator is removed.\\

\section{Effective Hamiltonian for $\phi^4$ scalar field theory}
The methods described above can be applied to field theory in $\nu$ spatial
dimensions. In this section  we discuss the one--component theory,
i.e. the $\phi^4$ model. The expectation value of the Hamiltonian with
respect to a Gaussian trial state is calculated, minimized and
subtracted from the Hamiltonian. This yields an effective Hamiltonian
containing a renormalized mass and new interaction terms. The latter only
appear if the original reflection symmetry, $ \phi \rightarrow - \phi$, is
spontaneously
broken.
In this method symmetry breaking effects are taken into account in
a nonperturbative way. The resulting effective Hamiltonian
may be treated in a perturbative way. In field theory the appearing
integrals are ultraviolet divergent; in the following they are assumed
to be regularized via a momentum cutoff or dimensional regularization.
One may introduce renormalization constants for the mass,
the coupling constant and the wave function. In
dimensional regularization one can avoid an explicit mass
renormalization. It is important to see that the effective
Hamiltonian scheme contains a new physical mass.
In this section we do not perform an explicit
renormalization. We merely assume the theory to be regularized and that
the resulting equations can be solved. Details of a full renormalization
and explicit solutions will be discussed in section (5).

The Hamiltonian for $\phi^4$ theory in terms of "bare"
quantities is given by
\begin{equation}
H = \int d^{\nu}x \lk[ \frac{1}{2} \pi^{2}_{B} +
\frac{1}{2}(\nabla\phi_{B})^2 +\frac{1}{2}m_{B}^{2}\phi_{B}^{2}
+\lambda_{B}\phi^{4}\rk],
\end{equation}
where $\pi_{B}$ is the conjugate momentum of $\phi_{B}$,
\begin{equation}
\lk[\pi_{B}(\vec{x}),\phi_{B}(\vec{y})\rk]=-i\delta^{\nu}(\vec{x}-\vec{y}).
\label{eq:commu}
\end{equation}
This commutator is to be understood as an equal time commutator. Since
we work in a time--independent scheme, the time variable is suppressed. We
explicitly introduce a field renormalization constant $Z_{\phi}$,
\begin{eqnarray}
 & \phi_{B}=Z_{\phi}^{\frac{1}{2}}\phi, & \nonumber \\
 & \pi_{B}=Z_{\phi}^{-\frac{1}{2}}\pi. &
\end{eqnarray}
The canonical commutator, eq. (\ref{eq:commu}) is preserved in this way.
Inserting this  into the Hamiltonian yields
\begin{equation}
H = \int d^{\nu}x \lk[ \frac{1}{2} Z_{\phi}^{-1}\pi^{2} +
\frac{1}{2}Z_{\phi}(\nabla\phi)^2 +\frac{1}{2}m_{B}^{2}Z_{\phi}\phi^{2}
+\lambda_{B}Z_{\phi}^{2}\phi^{4}\rk].
\end{equation}
The variational calculation starts with the introduction of a trial
ground state, $ \vacket_{\Omega,\phi_{0}}$. It is defined via
\begin{equation}
a_{\Omega}(\vec{k})\vacket_{\Omega,\phi_{0}}=0,
\end{equation}
with corresponding field (momenta) expansion
\begin{eqnarray}
  \phi(\vec{x}) &=& \phi_{0}+Z_{\phi}^{-\frac{1}{2}} \int (dk)_{\Omega}
  \lk[ a_{\Omega}(\vec{k})e^{i\vec{k} \cdot \vec{x}}+
  a_{\Omega}^{\dagger}(\vec{k}) e^{-i\vec{k} \cdot \vec{x}} \rk],\nonumber \\
  \pi(\vec{x})  &=& -iZ_{\phi}^{\frac{1}{2}} \int (dk)_{\Omega}
  \omega(\vec{k})
  \lk[ a_{\Omega}(\vec{k})e^{i\vec{k} \cdot \vec{x}}-
  a_{\Omega}^{\dagger}(\vec{k}) e^{-i\vec{k} \cdot \vec{x}} \rk].
\end{eqnarray}
The energies $\omega(\vec{k})=\sqrt{\vec{k}^2+\Omega^2}$ also appear in
the measure
\begin{equation}
  (dk)_{\Omega}=\frac{d^{\nu}k}{2(2\pi)^{\nu}\omega(\vec{k})},
\end{equation}
as well as in the canonical commutation relations of the creation and
annihilation operators
\begin{equation}
  \lk[a_{\Omega}(\vec{k}),a_{\Omega}^{\dagger}(\vec{k'})\rk]=
2(2\pi)^{\nu}\omega(\vec{k}) \delta^{\nu}(\vec{k}-\vec{k'}).
\end{equation}
Note that the $\phi_{0}$ dependences
of $a_{\Omega}(\vec{k})$ and $a_{\Omega}^{\dagger}(\vec{k'})$
are also suppressed. The state $ \vacket_{\Omega, \phi_{0}}$ is normalized,
\begin{equation}
 _{\Omega, \phi_{0}}\vacbra\vacket_{\Omega, \phi_{0}}=1.
\end{equation}
The quantities $\Omega$ and $\phi_{0}$ are the variational parameters of
the calculation.

As a next step we again normal order with respect to the
operators $a_{\Omega}$ and $a_{\Omega}^{\dagger}$;
this normal ordering is denoted
by $ : \; : $ . One obtains
\begin{eqnarray}
   : \phi^{2} :\ &=& \phi^{2}-Z_{\phi}^{-1} I_{0}(\Omega^2),   \nonumber \\
   : \phi^{4} :\ &=& \phi^{4}-6Z_{\phi}^{-1} I_{0}(\Omega^2):\phi^{2}:
                     -3Z_{\phi}^{-2}I_{0}(\Omega^2)^2,   \nonumber \\
   :(\nabla\phi)^{2} :\ &=& (\nabla\phi)^2-Z_{\phi}^{-1} I_{1}(\Omega^2)
                     +Z_{\phi}^{-1}\Omega^2 I_{0}(\Omega^2), \nonumber \\
   : \pi^{2} :\ &=& \pi^{2}-Z_{\phi} I_{1}(\Omega^2),
\label{eq:noro}
\end{eqnarray}
where the integrals $\IN$ are defined as
\begin{equation}
 \IN(\Omega^2)=\int
        (dk)_{\Omega}(\omega^{2}(\vec{k}))^N.
\end{equation}
For certain combinations of $\nu$ and $N$ these integrals are divergent. Here
we assume them to be regularized in such a way that the "naive" relation
\begin{equation}
\frac{d\IN(\Omega^2)}{d\Omega}=(2N-1)\Omega I_{\sssty N-1}(\Omega^2),
\end{equation}
holds for the regularized equations.
With eqs. (\ref{eq:noro}) one can write the Hamiltonian in normal
ordered form
\begin{eqnarray}
  H &=& : \int d^{\nu}x   \bigg[ \frac{1}{2} Z_{\phi}^{-1}\pi^{2} +
          \frac{1}{2}Z_{\phi}(\nabla\phi)^2 \nonumber \\
    & & +\ \frac{1}{2}(m_{B}^{2}+12\lambda_{B}I_{0})Z_{\phi}\phi^{2}
        +\lambda_{B}Z_{\phi}^{2}\phi^{4}+I_{1}-\frac{1}{2}\Omega^2 I_{0}
        +\frac{1}{2}m_{B}^{2}I_{0}+3\lambda_{B}I_{0}^2  \bigg]:,
\end{eqnarray}
where we suppressed the argument of the integrals: $\IN\equiv\INO$.
The expectation value
of the Hamiltonian density, ${\cal{H}}$, with respect to the trial state
is known as the Gaussian Effective Potential (GEP) \cite{Jackiw90}
\begin{eqnarray}
  V_{G}(\phi_{0},\Omega^2) &=&
 _{\Omega, \phi_{0}}\vacbra|\,{\cal{H}}\,\vacket_{\Omega, \phi_{0}}=\nonumber\\
  &=& I_{1}+\frac{1}{2}(m_{B}^{2}-\Omega^2)I_{0}+3\lambda_{B}I_{0}^2
+\frac{1}{2}(m_{B}^{2}+12\lambda_{B}I_{0})Z_{\phi}\phi_{0}^{2}
+\lambda_{B}Z_{\phi}^{2}\phi_{0}^{4}.
\label{eq:GEPO1}
\end{eqnarray}
Because of the possibility of spontaneous symmetry breaking,
normal ordered fields still can contribute to $V_{G}$, e.g.
\begin{equation}
_{\Omega, \phi_{0}}\vacbra|:\phi^2:\vacket_{\Omega, \phi_{0}}=\phi_{0}^{2}.
\end{equation}
The variational principle, which is expressed by the equations
$\frac{\partial V_{G}}{\partial\Omega}=0$
and $\frac{\partial V_{G}}{\partial\phi_{0}}=0$,
yields the gap equation
\begin{equation}
\Omega^2=m_{B}^{2}+12\lambda_{B}\lk[I_{0}+Z_{\phi}\phi_{0}^{2}\rk],
\label{eq:gap}
\end{equation}
and
\begin{equation}
m_{B}^{2}Z_{\phi}\phi_{0}+4\lambda_{B}Z_{\phi}^{2}\phi_{0}^{3}+
12\lambda_{B}I_{0}Z_{\phi}\phi_{0}=0.
\label{eq:minphi}
\end{equation}
It has to be verified that these equations do correspond
to a minimum \cite{Gasen}.
Here we proceed by assuming that these
equations hold and that their solutions indeed minimize the trial
vacuum energy. The vacuum expectation value is subtracted from the
Hamiltonian:
\bea
  H_{R} &=& H - \int d^{\nu}x V_{G} =
            : \int d^{\nu}x   \bigg[ \frac{1}{2} Z_{\phi}^{-1}\pi^{2} +
              \frac{1}{2}Z_{\phi}(\nabla\phi)^2 \nonumber\\
        & &   +\ \frac{1}{2}(m_{B}^{2}+12\lambda_{B}I_{0})Z_{\phi}
              (\phi^{2}-\phi_{0}^{2})
              +\lambda_{B}Z_{\phi}^{2}(\phi^{4}-\phi_{0}^{4})   \bigg]:.
\eea
The bare mass $m_{B}^{2}$ can be eliminated with the
gap equation, eq. (\ref{eq:gap}),
\bea
H_{R} &=&\
    : \int d^{\nu}x   \bigg[ \frac{1}{2} Z_{\phi}^{-1}\pi^{2} +
    \frac{1}{2}Z_{\phi}(\nabla\phi)^2
    +\frac{1}{2}\Omega^{2}Z_{\phi}(\phi^{2}-\phi_{0}^{2}) \nonumber\\
& & +\ \lambda_{B}Z_{\phi}^{2}(\phi^{4}-\phi_{0}^{4})
    -6\lambda_{B}Z_{\phi}^{2}\phi_{0}^{2}(\phi^{2}-\phi_{0}^{2})   \bigg]:,
\eea
and $H_R$ is rewritten in terms of the fluctuating field
$ \tilde{\phi}:=\phi-\phi_{0}$ ($\tilde{\pi}=\pi$),
\begin{equation}
H_{R}=\
: \int d^{\nu}x \lk[ \frac{1}{2} Z_{\phi}^{-1}\tilde{\pi}^{2} +
\frac{1}{2}Z_{\phi}(\nabla\tilde{\phi})^2
+\frac{1}{2}\Omega^{2}Z_{\phi}\tilde{\phi}^{2}
+\lambda_{B}Z_{\phi}^{2}\tilde{\phi}^{4}
+4\lambda_{B}Z_{\phi}^{2}\phi_{0}\tilde{\phi}^{3} \rk]:.
\end{equation}
This simple form, in particular the vanishing of the linear term
in $\tilde{\phi}$,
follows from the variational equations (\ref{eq:gap}, \ref{eq:minphi}).
Redefining  field and momentum as,
\be
  \tilde{\phi}=Z_{\phi}^{-\frac{1}{2}}\bar{\phi},\qquad
  \tilde{\pi}=Z_{\phi}^{\frac{1}{2}}\bar{\pi},
\ee
finally yields the effective Hamiltonian
\begin{equation}
H_{R}=\
: \int d^{\nu}x \lk[ \frac{1}{2} \bar{\pi}^{2} +
\frac{1}{2}(\nabla\bar{\phi})^2
+\frac{1}{2}\Omega^{2}\bar{\phi}^{2}
+\lambda_{B}\bar{\phi}^{4}
+4\lambda_{B}Z_{\phi}^{\frac{1}{2}}\phi_{0}\bar{\phi}^{3} \rk]:.
\end{equation}
Obviously, the mass is given by the variational parameter
$\Omega^2$. If the symmetry is spontaneously broken, i.e. $\phi_{0} \neq 0$,
the physical mass and the
vacuum expectation value are related by (cf. eq. (\ref{eq1:Omega}))
\begin{equation}
\Omega^2=8\lambda_{B}Z_{\phi}\phi_{0}^{2}.
\label{eq:OmePhi}
\end{equation}
A similar relation appears in the $O(2)$--model and has
important consequences for phenomenology.

Standard perturbation theory with the unperturbed free Hamiltonian $H_0$
\be
  H_0 =\ :\int d^{\nu}x\,
          \lk[\eh\bar{\pi}^2+\eh(\nab\bar{\phi})^2+\eh\Omq\bar{\phi}^2\rk]:
\ee
and the interaction term $H_I$
\be
  H_I =\ :\int d^{\nu}x\,
          \lk[4\lba Z_{\phi}^{\eh}\phio\bar{\phi}^3+\lba\bar{\phi}^4\rk]:
\ee
yields the ground state energy
\be
  \frac{1}{V}\vacbra|H\vacket = \VG(\phio)+
  \frac{1}{V}\sum_{n=1}^{\infty}\frac{(-i)^n}{n!}{\cal T}
  \int dt_1...\int dt_n
  \vacbra|H_I(0)H_I(t_1)...H_I(t_n)\vacket,
\ee
where $H_I(t)$ is the interaction Hamiltonian in the interaction picture:
\be
   H_I(t) := e^{iH_0t}H_Ie^{-iH_0t}.
\ee
In ref. \cite{Cea94} the expectation value $\la H\ra/V$ is called generalized
Gaussian Effective Potential.
It consists of the standard GEP, obtained variationally, and
perturbative corrections due to the effective interaction Hamiltonian.
Note that the  latter contains a new, cubic interaction term generated
nonperturbatively by spontaneous symmetry breaking.

\section{Effective Hamiltonian for the $O(2)$--model and Goldstone
bosons}
Two--component self--interacting scalar field theory
is the paradigm for the Goldstone mechanism. A simple, classical
treatment serves as textbook example for the occurrence of massless
particles due to the spontaneous breaking of a continuous symmetry. It
merely consists of a longitudinal shift of the field by a non--zero
vacuum expectation value which is found by minimization of the
classical potential. A vanishing mass term
with respect to the transverse direction is interpreted to correspond
to a massless excitation, i.e. the Goldstone boson.

Up to now, a satisfactory quantum mechanical analysis is lacking. Of
course, quantum corrections could destroy the simple, classical picture
and its consequences sketched above. In fact, this necessarily happens
in one space dimension because of Coleman's theorem \cite{Coleman66}.
Since exact,
analytical solutions, even restricted to the ground state properties and
masses, are beyond present possibilities, approximations have to
be made. A variational approach, using Gaussian wave functionals
with two mass parameters,
indeed yields spontaneous breaking of the $O(2)$ rotational symmetry
\cite{Stevenson87}. As mentioned in the introduction both excitations,
however, are massive; in conflict with Goldstone's theorem. The reason is that
the symmetry is not only broken because of a non--vanishing vacuum
expectation value of the field but also by the use of different masses
in the wave functional.
The variational ansatz was formulated in the "Cartesian"
basis; a formulation in radial and angular "coordinates" may avoid this
problem. Conceptual problems with the quantization have
prohibited such an analysis until now.

The (Cartesian) Gaussian ansatz with different masses is also not
compatible with an interpretation in terms of charged scalar particles.
This happens in the Abelian Higgs model: the Cartesian components of the
field are combined in a complex field, describing particle and
anti--particle excitations. Of course, particles and anti--particles
should have the same mass. The problems concerning the ansatz with
two mass parameters \cite{Stevenson87} will be discussed
in more detail in section (6).

In view of these arguments we do a variational calculation with one
mass. The procedure retraces the steps outlined in the introduction.
First, the $O(2)$--symmetry can only be spontaneously broken due to a
non--vanishing expectation value of the field. The second step, of course
absent in the symmetric phase, is to
diagonalize the quadratic part of the effective Hamiltonian.
In contrast to the one--component case this is, in
principle, a non--trivial Bogoliubov transformation. It is necessary
because of the use of only one mass parameter and, consequently,
only one gap equation. This transformation finally leads to
two different masses.
In fact, in the broken phase we obtain zero mass
excitations, i.e. the Goldstone bosons.

The Hamiltonian for the $O(2)$--model is given by
\begin{equation}
H = \int d^{\nu}x \lk[ \frac{1}{2} (\pi^{2}_{1B}+\pi^{2}_{2B}) +
\frac{1}{2}((\nabla\phi_{1B})^2+(\nabla\phi_{2B})^2)
+\frac{1}{2}m_{B}^{2}\phi_{B}^{2}
+\lambda_{B}\phi_{B}^{4}\rk],
\label{eq:HO2}
\end{equation}
where $\phi_{B}^2=\phi_{1B}^2+\phi_{2B}^2$. The equal time commutation
relations are
\begin{equation}
[\pi_{kB}(\vec{x}),\phi_{lB}(\vec{y})]=
-i\delta_{kl}\delta^{\nu}(\vec{x}-\vec{y}).
\label{eq:commu2}
\end{equation}
Again we introduce the renormalization constant $Z_{\phi}$ and rewrite
the Hamiltonian in terms of renormalized fields and momenta,
\begin{equation}
H = \int d^{\nu}x \lk[ \frac{1}{2}Z_{\phi}^{-1}(\pi^{2}_{1}+\pi^{2}_{2}) +
\frac{1}{2}Z_{\phi}((\nabla\phi_{1})^2+(\nabla\phi_{2})^2)
+\frac{1}{2}Z_{\phi}m_{B}^{2}\phi^{2}
+\lambda_{B}Z_{\phi}^2\phi^{4}\rk].
\end{equation}
Note that only one wave function renormalization $Z_{\phi}$ is necessary.
Analogous to the one--component theory we introduce a trial ground state
\begin{equation}
\vacket_{\Omega,\vec{\phi}_{0}}=
\vacket_{\Omega,\phi_{01}}\otimes\:\vacket_{\Omega,\phi_{02}},
\end{equation}
\begin{eqnarray}
  \phi_1(\vec{x}) &=& \phi_{01}+Z_{\phi}^{-\frac{1}{2}} \int (dk)_{\Omega}
  \lk[ a_{1}(\vec{k})e^{i\vec{k} \cdot \vec{x}}+
  a_{1}^{\dagger}(\vec{k}) e^{-i\vec{k} \cdot \vec{x}} \rk],\nonumber \\
  \phi_2(\vec{x}) &=& \phi_{02}+Z_{\phi}^{-\frac{1}{2}} \int (dk)_{\Omega}
  \lk[ a_{2}(\vec{k})e^{i\vec{k} \cdot \vec{x}}+
  a_{2}^{\dagger}(\vec{k}) e^{-i\vec{k} \cdot \vec{x}} \rk],\nonumber \\
  \pi_1(\vec{x})  &=& -iZ_{\phi}^{\frac{1}{2}} \int (dk)_{\Omega}
  \omega(\vec{k})
  \lk[ a_{1}(\vec{k})e^{i\vec{k} \cdot \vec{x}}-
  a_{1}^{\dagger}(\vec{k}) e^{-i\vec{k} \cdot \vec{x}} \rk],\nonumber\\
  \pi_2(\vec{x})  &=& -iZ_{\phi}^{\frac{1}{2}} \int (dk)_{\Omega}
  \omega(\vec{k})
  \lk[ a_{2}(\vec{k})e^{i\vec{k} \cdot \vec{x}}-
  a_{2}^{\dagger}(\vec{k}) e^{-i\vec{k} \cdot \vec{x}} \rk],
\label{eq:An2}
\end{eqnarray}
where the index $\Omega$ of the creation and annihilation operators
is omitted. We  normal order with respect to the trial
ground state state (cf. eqs. (\ref{eq:noro})):
\begin{eqnarray}
          : \phi_{j}^{2} : &=& \phi_{j}^{2}-Z_{\phi}^{-1}\InO,  \nonumber \\
              : \phi^{4} : &=& \phi^{4}-8Z_{\phi}^{-1}\InO:\phi^{2}:
                               -8Z_{\phi}^{-2}\InO^2,  \nonumber \\
   :(\nabla\phi_{j})^{2} : &=& (\nabla\phi_{j})^2-Z_{\phi}^{-1}\IeO
                               +Z_{\phi}^{-1}\Omega^2\InO,   \nonumber \\
           : \pi_{j}^{2} : &=& \pi_{j}^{2}-Z_{\phi}\IeO,
\label{eq:noro2}
\end{eqnarray}
where $j=1,2$.
In the following we again suppress the argument of the
regularized integrals: $\IN\equiv\INO$.
The Hamiltonian in normal ordered form reads
\begin{eqnarray}
  H &=& : \int d^{\nu}x   \bigg[ \frac{1}{2}
Z_{\phi}^{-1}(\pi_{1}^{2}+\pi_{2}^{2})
         + \frac{1}{2}Z_{\phi}((\nabla\phi_{1})^2+(\nabla\phi_{2})^2)
         \nonumber\\
    & & +\ \frac{1}{2}(m_{B}^{2}+16\lambda_{B}I_{0})Z_{\phi}\phi^{2}
        +\lambda_{B}Z_{\phi}^{2}\phi^{4}+2I_{1}
        +(m_{B}^{2}-\Omq)I_{0}+8\lambda_{B}I_{0}^2  \bigg]:.
\end{eqnarray}
The expectation value of the Hamiltonian density in the trial ground state is
readily calculated
from this expression,
\begin{equation}
 V_{G}(\phi_{0},\Omega^2)=
 \frac{1}{2}(m_{B}^{2}+16\lambda_{B}I_{0})Z_{\phi}\phi_{0}^{2}
+\lambda_{B}Z_{\phi}^{2}\phi_{0}^{4}
+2I_{1}+(m_{B}^{2}-\Omega^2)I_{0}+8\lambda_{B}I_{0}^2,
\label{eq:VGO2}
\end{equation}
with $\phi_{0}^{2}=\phi_{01}^{2}+\phi_{02}^{2}$.
The variational principle yields
\begin{equation}
\Omega^2=m_{B}^{2}+8\lambda_{B}\lk[2I_{0}+Z_{\phi}\phi_{0}^{2}\rk],
\label{eq:gap2}
\end{equation}
and
\begin{equation}
m_{B}^{2}Z_{\phi}\phi_{0}+4\lambda_{B}Z_{\phi}^{2}\phi_{0}^{3}+
16\lambda_{B}I_{0}Z_{\phi}\phi_{0}=0.
\label{eq:minphi2}
\end{equation}
For $\phioq\not=0$ these equations imply $\Omq=4\lba\phioq Z_{\phi}$.
Note that only the magnitude $\phi_{0}^{2}$ follows from these equations.
Because of
the rotational symmetry the direction is arbitrary. Furthermore, these
equations have to be supplemented by the condition that the
extremum indeed corresponds to a minimum. As in the previous section we
assume these equations to be fulfilled and continue by subtracting the
vacuum expectation value from the Hamiltonian
\begin{eqnarray}
 H_{R} &=& H - \int d^{\nu}x V_{G}=  \nonumber \\
       &=& :\int d^{\nu}x   \bigg[ \frac{1}{2}
           Z_{\phi}^{-1}(\pi_{1}^{2}+\pi_{2}^{2})+\frac{1}{2}Z_{\phi}
           ((\nabla\phi_{1})^2+(\nabla\phi_{2})^2)  \nonumber \\
       & & +\ \frac{1}{2}(m_{B}^{2}+16\lambda_{B}I_{0})Z_{\phi}
           (\phi^{2}-\phi_{0}^{2})
           +\lambda_{B}Z_{\phi}^{2}(\phi^{4}-\phi_{0}^{4})   \bigg]:.
\end{eqnarray}
The bare mass parameter is eliminated via the variational principle,
i.e. with eq. (\ref{eq:gap2}),
\begin{eqnarray}
 H_{R} &=& : \int d^{\nu}x   \bigg[ \frac{1}{2}
Z_{\phi}^{-1}(\pi_{1}^{2}+\pi_{2}^{2})
+ \frac{1}{2}Z_{\phi} ((\nabla\phi_{1})^2+(\nabla\phi_{2})^2)  \nonumber \\
& & +\ \frac{1}{2}(\Omega^{2}-8\lambda_{B}Z_{\phi}\phi_{0}^{2})
Z_{\phi}(\phi^{2}-\phi_{0}^{2})
+\lambda_{B}Z_{\phi}^{2}(\phi^{4}-\phi_{0}^{4})   \bigg]:.
\label{eq:Hsy}
\end{eqnarray}
Concomitantly, the divergent integrals have been removed.
For $\phi_{0}=0$ one again finds (after rescaling) that the
masses are both given by $\Omega^{2}$. The resulting effective
Hamiltonian obviously is $O(2)$ symmetric.

In the following we will consider the case $\phi_{0} \ne 0$.
The shifted fields $\tilde{\phi}$ are defined as
\begin{eqnarray}
& \tilde{\phi}_{1}=\phi_{1}-\phi_{01}=\phi_{1}-\phi_{0}\cos\alpha ,&
\nonumber \\
& \tilde{\phi}_{2}=\phi_{2}-\phi_{02}=\phi_{2}-\phi_{0}\sin\alpha .&
\label{eq:shift2}
\end{eqnarray}
Then one obtains
\begin{equation}
\phi^{2}-\phi_{0}^{2}=\tilde{\phi}^{2}+2\phi_{0}L(\tilde{\phi}),
\end{equation}
and
\begin{equation}
\phi^{4}-\phi_{0}^{4}=\tilde{\phi}^{4}+
4\phi_{0}L(\tilde{\phi})\tilde{\phi}^{2}+
4\phi_{0}{^2}L^2(\tilde{\phi})+2\tilde{\phi}^{2}\phi_{0}^{2}
+4\phi_{0}^{3}L(\tilde{\phi}),
\end{equation}
where
\begin{equation}
L(\tilde{\phi})=\tilde{\phi}_{1}\cos\alpha+\tilde{\phi}_{2}\sin\alpha.
\end{equation}
Substituting this in the effective Hamiltonian and using the extremum
conditions, eqs. (\ref{eq:gap2}, \ref{eq:minphi2}), gives
\begin{eqnarray}
     H_{R} &=&\ : \int d^{\nu}x   \bigg[ \frac{1}{2} Z_{\phi}^{-1}
     (\tilde{\pi}_{1}^{2}+\tilde{\pi}_{2}^{2})
     + \frac{1}{2}Z_{\phi} ((\nabla\tilde{\phi}_{1})^2
     +(\nabla\tilde{\phi}_{2})^2) \nonumber \\
& &  +\ \Omega^{2}Z_{\phi}L^{2}(\tilde{\phi})
     +\lambda_{B}Z_{\phi}^{2}(\tilde{\phi}^{4}+
     4\phi_{0}L(\tilde{\phi})\tilde{\phi}^{2})   \bigg]:.
\end{eqnarray}
In terms of the rescaled fields $ \bar{\phi}_{j} $ and momenta
$ \bar{\pi}_{j} $,
\begin{eqnarray}
 & \tilde{\phi}_{j}=Z_{\phi}^{-\frac{1}{2}}\bar{\phi}_{j}, & \nonumber \\
 & \tilde{\pi}_{j}=Z_{\phi}^{\frac{1}{2}}\bar{\pi}_{j}, &
\end{eqnarray}
one finally gets the effective Hamiltonian for the vacuum with spontaneously
broken symmetry. It is  characterized by $\phio$ and $\alpha$:
\bea
 H_{R} &=&\ : \int d^{\nu}x   \bigg[\frac{1}{2}
    (\bar{\pi}_{1}^{2}+\bar{\pi}_{2}^{2})
    + \frac{1}{2} ((\nabla\bar{\phi}_{1})^2
    +(\nabla\bar{\phi}_{2})^2) \nonumber\\
& & +\ \Omega^{2}L^{2}(\bar{\phi})
    +\lambda_{B}(\bar{\phi}^{4}+
    4Z_{\phi}^{\frac{1}{2}}\phi_{0}L(\bar{\phi})\bar{\phi}^{2})   \bigg]:.
\label{eq:Hsny}
\eea

We emphasize that spontaneous symmetry breaking, i.e.
the non--trivial vacuum and its consequences, are taken into
account in a nonperturbative way. Consider the quadratic part of the
Hamiltonian, which is normal ordered with respect to the operators
$a_{j}$. It is non--diagonal and needs to be diagonalized
by means of a Bogoliubov transformation. This can be done for an arbitrary
angle $\alpha$ as will be shown in Appendix A. In the following we will choose
most conveniently $\alpha=0$, for which the $\phi_{1}$ and $\phi_{2}$ modes
decouple and the mass term simplifies. The quadratic part of $H_R$ then has the
following form:
\be
   H_0 =\ :\int d^{\nu}x \lk[\frac{1}{2}
(\bar{\pi}_{1}^{2}+\bar{\pi}_{2}^{2})
+ \frac{1}{2} ((\nabla\bar{\phi}_{1})^2 +(\nabla\bar{\phi}_{2})^2)
+\eh(2\Omega^{2})\bar{\phi}_{1}^2\rk]:,
\ee
and corresponds to eq. (\ref{eq:Hnullqm})
for the quantum mechanical example.
In terms of the annihilation and creation operators one explicitly
has the non--diagonal form:
\begin{eqnarray}
  H_{0} &=&  \int (dk)   \bigg[ \omega(\vec{k})\lk[a_{1}^{\dagger}(\vec{k})
a_{1}(\vec{k})+a_{2}^{\dagger}(\vec{k})a_{2}(\vec{k})\rk]
  \nonumber \\
& & +\ \frac{\Omega^2}{4\omega(\vec{k})}  \bigg(\lk[
a_{1}^{\dagger}(\vec{k})a_{1}^{\dagger}(-\vec{k})+
2a_{1}^{\dagger}(\vec{k})a_{1}(\vec{k})+
a_{1}(\vec{k})a_{1}(-\vec{k})\rk]   \nonumber \\
& & -\ \lk[
a_{2}(\vec{k})a_{2}(-\vec{k})+
2a_{2}^{\dagger}(\vec{k})a_{2}(\vec{k})+
a_{2}^{\dagger}(\vec{k})a_{2}^{\dagger}(-\vec{k})\rk]  \bigg)   \bigg].
\end{eqnarray}
For the following it is more convenient to change
to a new set of creation and annihilation operators
with $\omega$--independent commutation relations
\begin{eqnarray}
a_{1}(\vec{k}) &=& \sqrt{2(2\pi)^{\nu}\omega(\vec{k})}a(\vec{k}), \nonumber \\
a_{2}(\vec{k}) &=& \sqrt{2(2\pi)^{\nu}\omega(\vec{k})}b(\vec{k}).
\end{eqnarray}
Analogous expressions hold for $a_1^{\dagger}$ and $a_2^{\dagger}$.
Recall that $\omega(\vec{k})=\sqrt{k^2+\Omega^2}$.
In order to diagonalize the quadratic part of the Hamiltonian we
define the following Bogoliubov transformation:
\bea
  u(\kv) &=& \cosh\alpha(\kv)a(\kv)+\sinh\alpha(\kv)\ad(-\kv) = U a(\kv)
U^{\dg}
           , \nonumber\\
  v(\kv) &=& \cosh\beta(\kv)b(\kv)+\sinh\beta(\kv)\bd(-\kv) = U b(\kv) U^{\dg},
\label{eq:Bogol}
\eea
with
\begin{equation}
   U = \exp\lk[-\int d^{\nu}k\lk\{
       \alpha(k)\lk(a^{\dg}(\kv)a^{\dg}(-\kv)-a(\kv)a(-\kv)\rk)+
         \beta(k)\lk(b^{\dg}(\kv)b^{\dg}(-\kv)-b(\kv)b(-\kv)\rk)\rk\}\rk] .
\label{eq:Bogol2}
\end{equation}
In terms of the operators $u$ and $v$
the quadratic part of the Hamiltonian $H_0$ now reads
\be
   H_0=\int d^{\nu}k\lk[\om_1(\kv)\ud(\kv)u(\kv)+
                           \om_2(\kv)\vd(\kv)v(\kv)\rk] +C,
\ee
where $C$ is an irrelevant c--number.
This determines
the functions $\alpha(k)$ and $\beta(k)$:
\bea
   \tanh\alpha(\kv) &=& \Om^{-2}\lk[2\om(\kv)^2+\Omq
                                -2\om(\kv)\sqrt{\kvq+2\Omq}\ \rk],\nonumber\\
   \tanh\beta(\kv) &=& \Om^{-2}\lk[-2\om(\kv)^2+\Omq
                                +2\om(\kv)k\rk].
\eea
In the limit $k\rar 0$ these "angles" agree with those of the quantum
mechanical example. The energy eigenvalues are readily obtained by
commuting $H_{0}$ with the quasiparticle operators $u$ and $v$
(cf. section (1)). This yields
\begin{eqnarray}
 & \om_1(\vec{k})=\sqrt{\vec{k}^2+M_{1}^2}, & \nonumber \\
 & \om_2(\vec{k})=\sqrt{\vec{k}^2+M_{2}^2}, &
\end{eqnarray}
where the physical masses are
\begin{eqnarray}
 & M_{1} = \Om\sqrt{2}, & \nonumber \\
 & M_{2} = 0 . &
\end{eqnarray}
The Goldstone boson is indeed massless. The ground state
$\vacket_{M_1, M_2}$ of the quadratic part of the Hamiltonian $H_0$
is given by
\be
\vacket_{M_1, M_2}= U \vacket_{\Omega, \vec{\phi}_{0}}.
\ee
It contains a coherent superposition of correlated boson
pairs with vanishing total momentum. Analogous states appear in
the theory of superfluidity and in the BCS theory.

Finally, we note that the explicit renormalization for
the one--component theory (cf. section (5)) can straightforwardly
be extended to the $O(2)$--model. Due to our ansatz with only one  mass
parameter, the structure of the Gaussian Effective Potentials,
eqs. (\ref{eq:GEPO1}) and (\ref{eq:VGO2}),
as well as the variational equations (cf. eqs. (\ref{eq:gap}),
(\ref{eq:minphi}), (\ref{eq:gap2}) and (\ref{eq:minphi2})) is identical.
These expressions only differ by some numerical
coefficients and we do not expect that the qualitative behaviour of
the solutions changes.

\section{Perturbation theory for the $O(2)$--model}
The Bogoliubov transformation, cf. eqs. (\ref{eq:Bogol}, \ref{eq:Bogol2}),
also modifies the non--quadratic terms of the effective Hamiltonian.
As we will show below, new quadratic terms are generated and thus the
Goldstone bosons could acquire a mass. These new terms, however, are
one--loop contributions and, in order to work consistently, all
one--loop contributions of the effective theory
must be taken into account. Alternatively
speaking, the possible mass corrections are of order $\lambda_{B}$ in the
effective theory and, again for consistency, one needs to include all
contributions of order $\lambda_{B}$. Note that, since $ \phi^{2}_{0}\sim
\frac{\Omega^2}{\lambda_{B}} $, $\phi_{0}\lambda_{B}$ is of order
$\sqrt{\lambda_{B}}$. In this way, the equivalence of the loop- and
coupling constant expansion is transparent to this order.

In this section we will explicitly show that the possible mass corrections
cancel and, as a consequence, that the Goldstone bosons remain massless in
field theory.
In order to control infrared divergences in perturbation theory
with massless particles we introduce an explicit
symmetry breaking term in the Hamiltonian. This also
slightly modifies the variational calculation. Finally, the
infrared regulator is taken to be zero.
Again we treat the quantum mechanical model first.
It is interesting that, although the one--loop contributions are
infrared finite, the zero frequency disappears in quantum mechanics.

\subsection{Quantum mechanics}
We start by adding an explicit symmetry breaking term to
the Hamiltonian:
\begin{equation}
H_{\delta} = -\delta^2 f x.
\end{equation}
The same variational approach, with the new variational parameters
$\Omega_{\delta} $ and $ x_{\delta} $ requires the minimization of
\begin{equation}
V_{G}^{\delta}=V_{G}-\delta^2 f \xdel.
\end{equation}
After doing this we take $f=x_{0}$, i.e. the expectation
value for $\delta=0$ (cf. section (1)),  and  we obtain
\begin{equation}
\Omega_{\delta}^{2}=4\lambda x_{\delta}^2 +\delta^2 +O(\delta^4).
\end{equation}
Furthermore, one can easily verify that
\begin{eqnarray}
  \Omega_{\delta}^{2} &=& \Omega^{2}+O(\delta^2) \nonumber\\
  \xdel^2 &=& x_{0}^2 + O(\delta^2).
\end{eqnarray}
More explicit expressions, which are irrelevant at the moment,
will be given for
the field theoretical case. The physical frequencies  correspond to the
diagonal quadratic part of the Hamiltonian
\begin{eqnarray}
  \omega_{1}^{2} &=& 2\Omega_{\delta}^{2}-\delta^2 +O(\delta^4), \nonumber\\
  \omega_{2}^{2} &=& \delta^2-O(\delta^4),
\end{eqnarray}
i.e.  the normal modes in the $y$--direction have
a finite frequency.
After the unitary Bogoliubov transformation
the Hamiltonian reads
\begin{equation}
H^{\delta}_{R}=H_{0}+H_I,
\end{equation}
with
\begin{equation}
H_{0}= \omega_{1} u^{\dagger} u + \omega_{2} v^{\dagger} v,
\end{equation}
and
\begin{eqnarray}
H_I &=& : \lambda \left[ \lk(\xtd^2+y^2\rk)^2
+ \lk(\frac{3}{\omega_{1}}+\frac{1}{\omega_{2}}-\frac{4}{\Omega_{\delta}}\rk)
\xtd^2
+
\lk(\frac{3}{\omega_{2}}+\frac{1}{\omega_{1}}-\frac{4}{\Omega_{\delta}}\rk)y^2
\right]
\nonumber \\
& & +\ 4\xdel\lambda \left[ \xtd y^2+
\lk(\frac{3}{2\omega_{1}}+\frac{1}{2\omega_{2}}
-\frac{2}{\Omega_{\delta}}\rk)\xtd+\xtd^3 \right] :_{u,v}.
\end{eqnarray}
As denoted the Hamiltonian is reordered with respect to the $(u,v)$
ground state. Moreover, a constant is omitted.

We are interested in the energy shift of the first excited state of the
$y$-oscillator. The energy difference between this state and the ground
state is $\delta$ for $H_0$, but will be changed by the perturbative
part of the Hamiltonian. We want to calculate its correction in
first order in $\lambda$.
The $y^2$ term, from now on denoted by $V_0$, gives a contribution
of order $\lambda$
and will be taken into account in first order perturbation theory.
The terms proportional to $\xdel$  must be
taken into account in second order perturbation theory since
$\xdel^2\lambda^2\sim\lambda$. Let us define
\be
U = 4\lambda\xdel(U_1+U_2+U_3),
\ee
where
\begin{eqnarray}
U_1 &=& :\xtd y^2: ,\nonumber \\
U_2 &=& \lk(\frac{3}{2\omega_{1}}+\frac{1}{2\omega_{1}}
-\frac{2}{\Omega_{\delta}}\rk)\xtd , \nonumber \\
U_3 &=& :\xtd^3:.
\end{eqnarray}
Here and in the following we omit the label $(u,v)$ at the normal
ordering symbol. The unperturbed energy eigenvalue is
\begin{equation}
E_{v}^{0}=\omega_{2}.
\end{equation}
The order $\lambda$ correction to $\Delta E_v$ is given by the energy
difference between the excited state $|\,n_x;n_y\ra=|\,0;1\ra$ and the ground
state $|\,n_x;n_y\ra=|\,0;0\ra$:
\begin{eqnarray}
\Delta E_{v} &=& \Delta E_{v}^{(1)}+ \Delta E_{v}^{(2)}   \nonumber \\
&=& \la 0;1\,|\,V_0\,|\,0;1\ra - \la 0;0\,|\,V_0\,|\,0;0\ra \nonumber \\
& & +\sum_{N} \frac{\la 0;1\,|\,U\,|\,N\ra\la N\,|\,U\,|\,0;1\ra}
{E_{v}^{0}-E_N}
- \sum_{N} \frac{\la 0;0|U|N\ra\la N|U|0;0\ra}{-E_N} .
\end{eqnarray}
The subtraction of the change in the ground state energy
corresponds to the omission
of disconnected contributions in field theory.
For $\Delta E_{v}^{(1)}$ one easily obtains
\begin{equation}
\Delta E_{v}^{(1)}= \frac{\lambda}{\omega_{1}}\lk(\frac{3}{\omega_{1}}
+\frac{1}{\omega_{1}} -\frac{4}{\Omega_{\delta}}\rk) .
\end{equation}
The calculation of the remaining part is
facilitated by noting that the possible contributing
intermediate states, starting from the state $|\,0;1\ra$, are
$ |\,1;1\ra , |\,1;3\ra $ via $U_1$, $ |\,1;1\ra $ via $U_2$, and $|\,3;1\ra$
via
$U_3$, respectively. Therefore, there is only one contributing
"interference" term, i.e. $U_1 U_2$ with the intermediate state $|\,1;1\ra$.
The $U_2^2$ and $U_3^2$ terms are obviously cancelled by the ground
state subtractions, whose relevant states are $ |\,1;2\ra, |\,1;0\ra$ and
$|\,3;0\ra$ for $U_1 , U_2$ and $U_3$, respectively. As a consequence, also
part of the $U_1^2$ contribution is cancelled.
In this way we obtain
\begin{eqnarray}
\Delta E_{v}^{(2)} &=& -\frac{16\lambda^2\xdel^2}
{2\,\omega_{2}^2\omega_{1}^2} \nonumber \\
& & -\ \frac{16\lambda^2\xdel^2}
{2\,\omega_{2}\omega_{1}^2(2\,\omega_{2}+\omega_{1})} \nonumber \\
& & -\ \frac{16\lambda^2\xdel^2}
{\omega_{2}\omega_{1}^2}\lk(\frac{3}{2\,\omega_{1}}
+\frac{1}{2\,\omega_{2}}-\frac{4}{\Omega_{\delta}}\rk),
\label{eq:dE2}
\end{eqnarray}
where the first line is the $U_1^2$ contribution via the intermediate
state $ |\,1;1\ra $.
The second line is also a $U_1^2$ contribution,
but from the intermediate state $|\,1;3\ra $ (including
the ground state subtraction). The third line is the contribution of
the interference term $U_1 U_2$.
Using $16\lambda^2\xdel^2=2\lambda(\omega_{1}^2-\omega_{2}^2)
+O(\delta^4)$ gives
\begin{eqnarray}
\Delta E_{v}^{(2)} &=& -\frac{\lambda}{\omega_{2}^2}+
\frac{\lambda}{\omega_{1}^2} \nonumber \\
& & +\ \lambda\frac{\omega_{2}^2-\omega_{1}^2}
{\omega_{1}\omega_{2}^2(2\,\omega_{2}+\omega_{1})} \nonumber \\
& & +\ \frac{\lambda}{\omega_{1}^2}-\frac{\lambda}{\omega_{2}}
\lk(\frac{3}{\omega_{1}}+\frac{1}{\omega_{2}}-\frac{4}{\Omega_{\delta}}\rk)
+O(\delta).
\end{eqnarray}
Further expanding in
$\delta$ and adding the terms finally yields
\begin{equation}
\Delta E_{v} = -\frac{\lambda}{\omega_{1}^2}+O(\delta).
\end{equation}
Consequently, all the possible infrared divergences cancel and the limit
$\delta \rightarrow 0$ is trivial. The (would be) zero frequency
becomes finite in perturbation theory.

Naively, one may expect more severe infrared problems in the
corrections to the frequency of the $x$--oscillator. An explicit
calculation, however, shows that also in this case these
infrared divergences cancel and the order $\lambda$ correction
is finite in the limit $\delta \rightarrow 0$. Of course, this does
not exclude infrared infinities in higher order.

\subsection{Field Theory}
The phenomenon we described above could generate a Goldstone
mass in field theory. We will explicitly demonstrate, however,
that due to a different infrared behaviour
this does not happen. In contrast to quantum mechanics the zero Goldstone mass
does not get changed in perturbation theory to this order.

We add an explicit symmetry breaking term to the Hamiltonian of
the $ O(2)$--model (cf. eq. (\ref{eq:HO2})),
\begin{equation}
H_{\delta}=H-f\delta^2\sqrt{Z_{\phi}}
\int d^{\nu}x \phi_{1}(\vec{x}) .
\end{equation}
We repeat the variational calculation of section (3) but with the
new parameters $\Omega_{\delta}$ and $\phi_{\delta}$. Consistent with the added
symmetry breaking term
we take $\alpha=0$ from the beginning, i.e. $\phi_{01}=\phi_{0}$
and $\phi_{02}=0$. The expectation value of the Hamiltonian density
now reads
\begin{equation}
V_{G}^{\delta}(\phi_{\delta},\Omega_{\delta})=
V_{G}(\phi_{\delta},\Omega_{\delta})-f\delta^2\sqrt{Z_{\phi}}
\phi_{\delta},
\end{equation}
with $V_G$ from the previous $O(2)$ calculation, eq. (\ref{eq:VGO2}).
Minimization
yields the same gap equation (\ref{eq:gap2}), but in the new variables.
The $\phi_{\delta}$ equation, however, is modified
\begin{equation}
Z_{\phi}\phi_{\delta}(m_{B}^{2}+
4\lambda_{B}Z_{\phi}\phi_{\delta}^{2}+
16\lambda_{B}I_{0})=\sqrt{Z_{\phi}}f\delta^2.
\label{eq:minphidel}
\end{equation}
For convenience we fix $f$ in the usual way, i.e. $f=\sqrt{Z_{\phi}}\phi_{0}$,
where $\phi_{0}$ is determined by eqs. (\ref{eq:gap2}) and (\ref{eq:minphi2}).
(These equations
also yield $\Omega$). With this choice one obtains
\begin{eqnarray}
\Omega_{\delta}^2 &=& \Omega^2+\delta^2\frac{2}
{1-8\lambda_{B}I_{-1}(\Omega^2)}+O(\delta^4), \nonumber \\
\phi_{\delta}^{2}&=&\phi_{0}^{2}+\frac{\delta^2}{4\lambda_{B}Z_{\phi}}
\frac{1+8\lambda_{B}I_{-1}(\Omega^2)}{1-8\lambda_{B}I_{-1}(\Omega^2)}
+O(\delta^4).
\end{eqnarray}
In the following we proceed as in section (3). First, the vacuum expectation
value of the Hamiltonian is subtracted and the bare mass is eliminated
with the gap equation. Secondly, we shift the field $\phi_{1} ,
\tilde\phi_{1}=\phi_{1}-\phi_{\delta}$ and insert eq. (\ref{eq:minphidel})
in order to
get rid of the linear terms. Finally, we rescale the fields and find
\begin{eqnarray}
 H_{R}^{\delta} &=&\ : \int d^{\nu}x   \bigg[\frac{1}{2}
    (\bar{\pi}_{1}^{2}+\bar{\pi}_{2}^{2})
    + \frac{1}{2} ((\nabla\bar{\phi}_{1})^2
    +(\nabla\bar{\phi}_{2})^2) \nonumber\\
& & +\frac{1}{2}M_{1}^{2}\bar{\phi}_{1}^2+\frac{1}{2}M_{2}^{2}\bar{\phi}_{2}^2
    +\lambda_{B}(\bar{\phi}^{4}+
    4Z_{\phi}^{-\frac{1}{2}}\phi_{0}\bar{\phi}_{1}\bar{\phi}^{2})   \bigg]:,
\end{eqnarray}
with
\begin{eqnarray}
M_{1}^{2}&=& 2 \Omega_{\delta}^2-\delta^2+O(\delta^4), \nonumber \\
M_{2}^{2}&=& \delta^2+O(\delta^4).
\end{eqnarray}
The normal ordering is still with respect to $\Omega_{\delta}$. The
Bogoliubov transformation which diagonalizes the quadratic part of
the Hamiltonian corresponds to a reordering with respect to  the
corresponding operators $u$ and $v$.
As before, this simplicity is due to the choice $\alpha=0$.
Apart from a constant, the result is
\begin{equation}
H_{R}^{\delta}=H_{0}+H_I,
\end{equation}
with the free, diagonal Hamiltonian
\begin{equation}
 H_{0} =\ : \int d^{\nu}x \lk[\frac{1}{2}
    (\bar{\pi}_{1}^{2}+\bar{\pi}_{2}^{2})
    + \frac{1}{2} ((\nabla\bar{\phi}_{1})^2
    +(\nabla\bar{\phi}_{2})^2)
+ \frac{1}{2}M_{1}^{2}\bar{\phi}_{1}^2+\frac{1}{2}M_{2}^{2}\bar{\phi}_{2}^2
\rk] :_{u, v}
\end{equation}
and the interaction terms
\begin{eqnarray}
H_I &=&\ : \int d^{\nu}x \lambda_{B}  \bigg[
    \bar{\phi}^{4}+
    4\sqrt{Z_{\phi}}\phi_{0}\bar{\phi}_{1}\bar{\phi}^{2}
\nonumber \\
& & +\ 4\sqrt{Z_{\phi}}\phi_{0}
\lk(3I_{0}(M_{1}^{2})+I_{0}(M_{2}^{2})-4I_{0}(\Omega_{\delta}^{2})\rk)
\bar{\phi}_{1}
\nonumber \\
& & +\ \lk(6I_{0}(M_{1}^{2})+2I_{0}(M_{2}^{2})-8I_{0}(\Omega_{\delta}^{2})\rk)
\bar{\phi}_{1}^2
\nonumber \\
& & +\ \lk(6I_{0}(M_{2}^{2})+2I_{0}(M_{1}^{2})-8I_{0}(\Omega_{\delta}^{2})\rk)
\bar{\phi}_{2}^2
  \bigg]:_{u, v}.
\end{eqnarray}
$H_{0}$ describes free particles with energies
$E_{1}^{(0)}(\vec{p})=\sqrt{M_{1}^2+\pvq}$
and $E_{2}^{(0)}(\vec{p})=\sqrt{M_{2}^2+\pvq}$,
for given 3--momentum $\vec{p}$. We are interested in possible
mass corrections to the Goldstone boson, i.e. particle 2. It is
convenient to take the limit $\vec{p} \rightarrow 0$ and
eventually we want to study the limit $\delta \downarrow 0$
in order to see whether the Goldstone boson remains massless.

The perturbative calculation is completely analogous to the
quantum mechanical example. The latter is actually contained in
the field theoretical framework; it formally corresponds to zero space
dimension ($\nu=0$) (without the particle interpretation).
The extension of the definitions of the potentials is straightforward
and will be assumed from now on. Again the $U_3$ pieces are disconnected
and, consequently, irrelevant. As in quantum mechanics, the first
correction is trivially calculated.
\begin{eqnarray}
  \Delta E_{2}^{(1)}(\vec{p}=0) &=&  \lim_{\vec{p} \rightarrow 0}
\frac{\la 0;\vec{p}\,|\,V_0\,|\,0;\vec{p}\ra} {\la
0;\vec{p}\,|\,0;\vec{p}\ra}\nonumber\\
&=& \frac{2\lambda_{B}}{M_2}\left[I_{0}(M_{1}^2)+3I_{0}(M_{2}^2)
-4I_{0}(\Omega_{\delta}^2)\right].
\end{eqnarray}
The rest of the calculation is slightly more involved than in
quantum mechanics since one has to integrate over the momenta
of the particles in the intermediate states. After subtracting
the disconnected pieces we obtain
\begin{eqnarray}
  \Delta E_{2}^{(2)}(\vec{p}=0) &=&
\frac{16\lambda_{B}^2Z_{\phi}\phi_{\delta}^2}{M_2}
\int\frac{d^{\nu}q}{2(2\pi)^3}\frac{1}{\sqrt{q^2+M_1^2}}
\frac{1}{\sqrt{q^2+M_2^2}} \frac{1}{M_2-\sqrt{q^2+M_1^2}-\sqrt{q^2+M_2^2}}
\nonumber \\
&-&
\frac{16\lambda_{B}^2Z_{\phi}\phi_{\delta}^2}{M_2}
\int\frac{d^{\nu}q}{2(2\pi)^3}\frac{1}{\sqrt{q^2+M_1^2}}
\frac{1}{\sqrt{q^2+M_2^2}} \frac{1}{M_2+\sqrt{q^2+M_1^2}+\sqrt{q^2+M_2^2}}
\nonumber \\
&-&
\frac{16\lambda_{B}^2Z_{\phi}\phi_{\delta}^2}{M_2}
\frac{1}{M_1^2}\left[3I_0(M_1^2)+I_0(M_2^2)-4I_0(\Omega_{\delta}^2)\right].
\label{eq:dE2FT}
\end{eqnarray}
Here the same order
of presentation as in eq. (\ref{eq:dE2}) has been chosen. Recall that some
ultraviolet regularization is assumed.
The numerator of the common prefactor can be written as
\begin{eqnarray}
16\lambda_{B}^2 Z_{\phi}\phi_{\delta}^2 &=& 4\lambda_{B}
(\Omega_{\delta}^2-\delta^2) + O(\delta^4) \nonumber \\
&=& 2\lambda_{B}(M_1^2-M_2^2) + O(\delta^4).
\end{eqnarray}
This enables us to rewrite eq. (\ref{eq:dE2FT}) as
\begin{eqnarray}
  \Delta E_{2}^{(2)}(\vec{p}=0) &=&
\frac{2\lambda_{B}}{M_2}   \bigg[
I_{0}(M_{1}^{2}) - I_{0}(M_{2}^{2})
\nonumber \\
& &+\ I_0(M_1^2) - I_0(M_2^2)
\nonumber \\
& &-\ 3I_0(M_1^2) - I_0(M_2^2) + 4I_0(\Omega_{\delta}^2)   \bigg]
+ O(\delta).
\label{eq:EFTF}
\end{eqnarray}
This equation is proved in Appendix B where also the quantum mechanical
result will be rederived.
Consequently,
\begin{equation}
\Delta E_{2}(\vec{p}=0)=\Delta E_{2}^{(1)}(\vec{p}=0)+
\Delta E_{2}^{(2)}(\vec{p}=0) =  O(\delta),
\end{equation}
which means that in the limit $\delta \downarrow 0$ the Goldstone
boson indeed remains massless to this order.
The $O(\delta)$ in the last two equations is only valid for
space--dimension $\nu \ge 2$, which is indeed relevant for the Goldstone
boson interpretation.
In the one--dimensional case a logarithm appears, i.e. one finds
$ O(\delta\ln\delta)$. This also vanishes in the limit
$\delta \downarrow 0$ and, consequently, does not reflect
Coleman's theorem \cite{Coleman66}.
In quantum mechanics there are finite corrections: the
frequency is non--zero even in the limit $\delta \downarrow 0$.
Again the non--normalizable plane waves (with zero momentum) are
avoided. Finally, just as in quantum mechanics,
the analogous calculation in field theory yields
that the one--loop corrections to
the massive excitation are infrared finite.

\section{Renormalization of the GEP at finite momentum cutoff}
Much effort has been put into the renormalization of the Gaussian Effective
Potential of scalar $\lpv$ \cite{Stevenson84,Consoli92,Stevenson87}.
Generally, it is believed  that after the regulator is removed, $\lpv$--theory
becomes trivial (see e.g. \cite{Luescher88}). We do not consider here the so
called "autonomous" renormalization scheme, which has been developed in ref.
\cite{Stevenson87} and critically
examined in  ref. \cite{Soto89}.
The great success of the Standard Model gives special importance to a detailed
understanding of spontaneous symmetry breaking  and of the Higgs--Mechanism.
One expects that the effective potential of the true underlying theory develops
a nontrivial minimum for certain values of the parameters.
The $\lpv$ theory might not be a sufficiently good approximation to the
underlying theory at energies much larger than some maximum value $\La$, which
may be given by the scale of Grand Unification or the breaking of
supersymmetry.
Therefore we explicitly keep a finite momentum cutoff, i.e. all modes with
momentum larger than $\La$ will be discarded and the regulator will not be
removed after the reparametrization in terms of the new "renormalized"
parameters. All dimensionful parameters, variables and resulting expressions
for the GEP will be expressed in units of the cutoff $\La$:
\bea
      x := \,\Omq/\Laq,&\qquad\qquad\quad \xn\!\! &:= \Omnq/\Laq,  \nonumber\\
   \mhq := m^2/\Laq,& \qquad\qquad\quad\Phioq\!\! &:= \phioq/\Laq,\nonumber\\
  \mrhq := \mrq/\Laq,\!& \qquad        \cVGPx\!\! &:= \VGpO/\La^4,
\label{4dimlos}
\eea
where $\Omnq$ is defined as the solution of the gap equation (\ref{eq:gap}) for
$\phi_0=0$. The effects of vacuum fluctuations become finite and the resulting
$\IN$--integrals can be written in units of $\La$ (see Appendix C).\\
In analogy to ref. \cite{Luescher88} we assert that the physical parameters do
not exceed one half of the value $\La$:
\be
  \sqrt{x}\le\eh,\qquad\Phio\le\eh.
\label{4Skaling}
\ee
As we will see later, the phenomenological results of the calculation are
rather sensitive to the exact value of this upper limit. In any case it should
not exceed unity.

\begin{figure}[p]
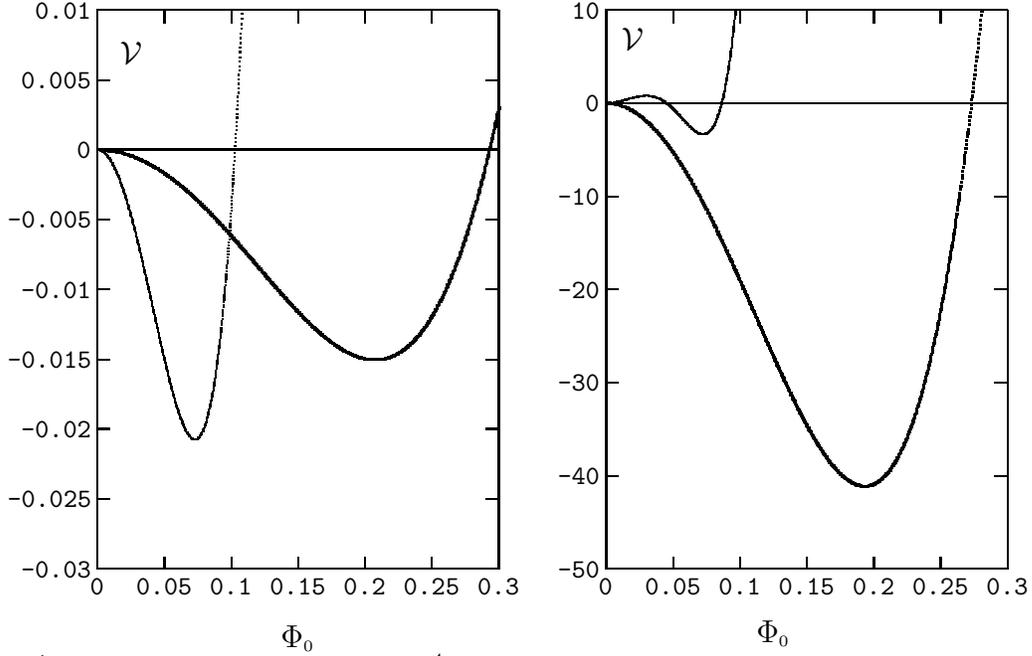

\begin{center}
\input{abbP1a} \input{abbP1b}\par\ \\
\caption[Das renormierte GEP]{\label{AB-GEP-cutmrqgeN}\fns The renormalized GEP
$\cVG\times 10^4$ (thin line) compared to the bare potential $\cV_{bare}\times
10^2$ (thick line). On the left side a Higgs--mass $\MH=200$ GeV has been
chosen, on the right side $\MH=1.2$ TeV. The bare potential has been rescaled
by a factor $100$.\\}
\end{center}
\end{figure}

We impose the renormalization conditions:
\bea
     \mrhq &:=&\ \ \ \left.\frac{d^2\cVG(\Phio)}{d\Phioq}\right |_{\Phioq=0},
       \label{4mre-Def}\\
     \lre &:=&
          \left.\frac{1}{4!}\frac{d^4\cVG(\Phio)}{d\Phio^4}\right|_{\Phioq=0}.
       \label{4lre-Def}
\eea
By inserting the expression for the GEP (eq. (\ref{eq:GEPO1})) we calculate the
flows of the bare--parameters $\mbhq, \lba$, i.e. their dependence on the
renormalized parameters $\mrhq, \lre$. In the self--consistent
Hartree--approximation we take $Z_{\phi}=1$ (see the discussion in ref.
\cite{Soto89}).\\
One has to distinguish between two cases, $\mrhq\ge 0$ and $\mrhq<0$:

a) $\mrhq\ge 0$:
The renormalization conditions together with the expression of the GEP lead to
\bea
    \mrhq &=& \mbhq +12\lba\In(\xn), \label{4MRHQ}\\
     \lre &=& \lba\frac{1-12\lba\Imi(\xn)}{1+6\lba\Imi(\xn)}.\label{4LRE}
\eea
Now a specific choice for $\xn$ has to be made. There are two possibilities for
the extremum condition
\be 0 = \frac{\ptd\cVG}{\ptd\sqrt{x}}
      =
\eh\,\sqrt{x}\,\Imi(x)\left(x-\mbhq-12\lba\left[\Phioq+\In(x)\right]\right),
\label{4ExtrBed}
\ee
to be satisfied. The first one is $x\equiv 0$, independent of $\Phio$ and the
bare parameters. The other is given by the solution of the "optimization
equation":\
\be
   x = \mbhq + 12\lba[\Phioq+\In(x)].\label{4OPT}
\ee
In the present case ($\mrhq\ge 0$) only
\be
   \xn = \mbhq + 12\lba\In(\xn) = \mrhq
\ee
applies (cf. Appendix D), where we used eq. (\ref{4MRHQ}) for the second
identity. By solving eq. (\ref{4MRHQ}) and eq. (\ref{4LRE}) one now obtains
expressions for $\lba, \mbhq$ in terms of the renormalized parameters. For the
explicit calculation of the renormalized GEP we use the parametrization
\bea\mbhq &=& \frac{\mhq}{\Imi(\xn)}-12\lba\In(0),\label{4mba-Fl}\\
     \lba &=& \frac{\eta}{\Imi(\xn)}\label{4lba-Fl},
\eea
which allows us to write the following equations in a more convenient form.
We will need a subtraction formula for the $\IN$--integrals,
\be
   \Inxn-\Inn = -\xh\Imixn+2f'(x), \label{4Subfor}
\ee
where $\xn=x(\Phio=0)$ has been chosen as a reference point.
The function $\fhx$ and its derivative $\fhxe$ are defined and
calculated in Appendix C.
Moreover, it is more transparent to start from the derivative of the GEP with
respect to $\Phioq$, to insert the flow eqs. (\ref{4mba-Fl}), (\ref{4lba-Fl})
as well as the optimization equation (\ref{4OPT}) and to integrate the
resulting expression:
\bea
     \frac{d\cVG}{d\Phioq}
      &=& \frac{1}{2\Phio}\frac{d\cVG}{d\Phio} =
          \eh\left(x-8\lba\Phioq\right) \nonumber\\
      &=& \frac{1}{\eseI}\left[\frac{\mhq}{2}+\zwe\fhxe+
          2\eta(1-\zwe)\Phioq\right],
\label{4dVGdPhioq}
\eea
where
\bea
     x &=& \frac{\mhq}{\Imi}+\frac{12\eta}{\Imi}
            \left[\Phioq-\xh\Imi+2\fhxe\right] \nonumber\\
      &=& \frac{1}{\eseI}\left[\mhq+\zwe\left(\Phioq+2\fhxe\right)\right].
\label{4Omgl}
\eea
Starting with eq. (\ref{4Omgl}) we omit the argument of $\Imixn$:
$\Imi\equiv\Imixn$).
Now expression (\ref{4dVGdPhioq}) is transformed into a total derivative via
\bea
     \dxdPhioq &=& \frac{\zwe}{\eseI}\left[1+2\dxdPhioq\fhxz\right]
                   \nonumber\\
      &=& \left[\frac{\eseI}{\zwe}-2\fhxz\right]^{-1}.
\label{4dxdPhioq}
\eea
Inserting this into eq. (\ref{4dVGdPhioq}) leads to
\be
    \frac{d\cVG}{d\Phioq} = \frac{\mhq+4\eta\emze\Phioq}{2\eseI} +
    \left(1-\frac{24\eta\fhxz}{\eseI}\right)\fhxe\dxdPhioq,
\label{4dVGdPq}
\ee
and after integration to
\bea
  \cVGP &=& \int_0^{\Phioq}\frac{d\cVG}{d\PhioSq}d\PhioSq=\nonumber\\
        &=& \fhx-\fhxn+\frac{\mhq\Phioq+2\eta\emze\Phio^4-
             24\eta\left(\fhxe^2-\fhxne^2\right)}{2\eseI}\nonumber\\
        &=& \eh\mrhq\Phioq+\lre\Phio^4+\fhx-\fhxn \nonumber\\
        & & \ \ -\frac{12\eta}{\eseI}
             \left(\Phioq\fhxne+\fhxe^2-\fhxne^2\right).
\eea
This is the resulting (dimensionless) expression for the renormalized GEP which
will be analyzed below with respect to the position of local or absolute minima
in terms of the renormalized parameters. Fig. 1 (right) shows an example
of the renormalized GEP for  $\mrhq>0$.

b) $\mrhq<0$:
In this case a minimization of the GEP at $\Phio=0$ is only possible through
$\xn=0$ (cf. eq. (\ref{4ExtrBed})): By definition there is no solution $\xn<0$
to the optimization equation (\ref{4OPT})). Thus eqs. (\ref{4mre-Def}),
(\ref{4lre-Def}) become
\bea \mrhq &=& \mbhq +12\lba\Inn, \\
      \lre &=& \lba,
\eea
and for the flows one obtains
\bea \mbhq &=& \mrhq - 12\lre\In(0),\label{4mbq-cl}\\
      \lba &=& \lre.\label{4lba-cl}
\eea
With this flow the optimization equation (cf. eq. (\ref{4OPT}))
\be
  x=\mrhq+12\lba\left[\Phioq+\In(x)-\Inxn\right]\label{4OmGl-mrq}
\ee
has no positive solution for $x$ within the interval
\be
  0\le\Phioq < \Phioqcr:=-\frac{\mrhq}{12\lba}. \label{4cl-Bereich}
\ee
Thus the integration has to be split into two parts:
\be
  \cVG(\Phioq<\Phioqcr) = \eh\mrhq\Phioq+\lre\Phio^4 \label{4GEP-cl},
\ee
\be
   \cVG(\Phioq\ge\Phioqcr) = \eh\mrhq\Phioqcr+\lre\Phiocr^4+
            \int_{\Phioqcr}^{\Phioq}\frac{d\cVG}{d\PhioSq}d\PhioSq.
\label{4GEP-gePhioqcr}
\ee
Fig. \ref{AB-GEP-cutmrqgeN} (left) shows an example for the renormalized GEP
for $\mrhq<0$.
One should note the difference in the scales of the bare and the renormalized
potentials.

The "scaling condition", eq. (\ref{4Skaling}), restricts the physical
observables with respect to the cutoff. In our
framework the self--consistent mass is given by the value of $\Om$ at the
absolute minimum of the GEP. This allows to express the upper limit for the
mass $\Omega$ in the broken phase in terms of
$v$, i.e. the vacuum expectation value of the scalar field $\Phi$ at the
minimum of the GEP (cf. eq. (\ref{eq:OmePhi})):
\be
  \Omq = \frac{8\eta v^2}{\Imixn}\label{4xv-vhq}.
\label{Omebou}
\ee
The second derivative of the GEP with respect to $\Phio$ has to be positive at
$\Phio=\vh$. This gives an upper bound on $\eta$:
\be
    0\le\eta\le\frac{1}{12}\frac{\Imixn}{\Imixv}.
\label{4eta-Absch}
\ee
The lower bound results from eq. (\ref{Omebou}).
Eqs. (\ref{Omebou}, \ref{4eta-Absch}) have to be consistent with
the upper limit for $\sqrt{x_v}<1/2$ (cf. eq. (\ref{4Skaling})); this yields
\be
  \Omq = \frac{8\eta v^2}{\Imixn}\le\frac{2v^2}{3\Imixv}\lsim(6,9\,v)^2.
  \label{4Omv-Obergr}
\ee
If one assumes that $\Om$ represents an approximation to the Higgs--mass, the
mass of the radial mode in the $U(1)$--model, then one obtains as an upper
limit to the Higgs--mass:
\be
  \MH \le 6.9\cdot246\,\mtext{GeV} \simeq 1.7\,\mtext{TeV}.
\ee
Here we use $v=(\sqrt{2}G_F)^{-1/2}\simeq 246$ GeV. This mass in turn
corresponds to a minimum cutoff of $3.4$ TeV. In comparison, the customary
upper limit to the Higgs--mass in the literature \cite{Gunion90} is $\MH \le
1.2\,\mtext{TeV}$,
obtained from unitarity arguments in $W^\pm$--scattering with Higgs exchange.
Of course, also our limit on $\MH$ may change when we consider the
complete Higgs model with vector fields.

On the basis of our upper bound we calculate the parameters which appear in our
renormalization procedure for a cutoff value of $3.4$ TeV, which is twice the
limit on the Higgs--mass. Furthermore, we choose three different values for the
Higgs--mass ($200$ GeV, $800$ GeV and $1.2$ TeV) and take the vacuum
expectation value to be $246$ GeV as above. The results are shown in Table 1.
As can be seen, the light Higgs bosons are coupled more weakly than the heavy
ones. The first two Higgs bosons correspond to case b) $\mrhq<0$, i.e. the
renormalized GEP has only one minimum at $v=246$ GeV and a local maximum at
$\Phio=0$. This also leads to $\lre$ being equal to $\lba$. The heavy Higgs
boson corresponds to case a) $\mrhq\ge 0$, so the renormalized potential has
two local minima, one at $\Phio=0$ as well as an absolute minimum at a finite
value of $\Phio$. One can show that case a) can only become relevant for cutoff
values larger than about $2.7$ TeV.
\begin{table}[tb]
\begin{center}
\begin{tabular}{|l|r|r|r|}
\hline
$\MH$   & $200$ GeV        & $800$ GeV         & $1200$ GeV         \\ \hline
\hline
$\lba$  & $0.083$          & $1.322$           & $2.974$            \\ \hline
$\lre$  & $0.083$          & $1.322$           & $-1.700$           \\ \hline
$\mbq$  & $-(405$ GeV$)^2$ & $-(1559$ GeV$)^2$ & $-(2261$ GeV$)^2$  \\ \hline
$\mrq$  & $-(136$ GeV$)^2$ & $-(326$ GeV$)^2$  & $(220$ GeV$)^2$    \\ \hline
$\Omnq$ & $0$              & $0$               & $(220$ GeV$)^2$    \\ \hline
\end{tabular}\\ \ \\
Table 1: {\fns Bare and renormalized couplings for three different
Higgs--masses\\ for a cutoff of $\La=3.4$ TeV and a vacuum expectation value
$v=246$ GeV.}\\
\end{center}
\end{table}
Figure \ref{AB-GEP-cutmrqgeN} shows the bare potential as well as the
renormalized GEP for Higgs--masses of $200$ GeV (left) and $1.2$ TeV (right).
The bare potential has been rescaled by a factor $100$.

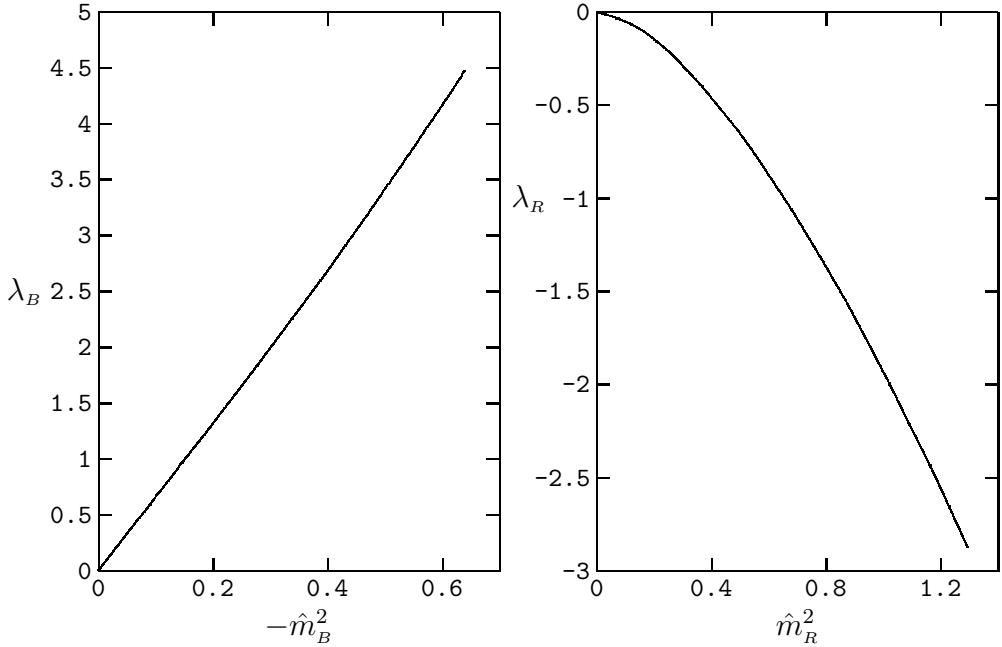
\begin{figure}[p]
\begin{center}
\setlength{\unitlength}{0.240900pt}
\ifx\plotpoint\undefined\newsavebox{\plotpoint}\fi
\begin{picture}(750,900)(0,0)
\font\gnuplot=cmtt10 at 10pt
\gnuplot
\sbox{\plotpoint}{\rule[-0.200pt]{0.400pt}{0.400pt}}%
\put(80.0,16.0){\rule[-0.200pt]{151.767pt}{0.400pt}}
\put(80.0,16.0){\rule[-0.200pt]{0.400pt}{211.510pt}}
\put(80.0,16.0){\rule[-0.200pt]{4.818pt}{0.400pt}}
\put(70,16){\makebox(0,0)[r]{0}}
\put(690.0,16.0){\rule[-0.200pt]{4.818pt}{0.400pt}}
\put(80.0,104.0){\rule[-0.200pt]{4.818pt}{0.400pt}}
\put(70,104){\makebox(0,0)[r]{0.5}}
\put(690.0,104.0){\rule[-0.200pt]{4.818pt}{0.400pt}}
\put(80.0,192.0){\rule[-0.200pt]{4.818pt}{0.400pt}}
\put(70,192){\makebox(0,0)[r]{1}}
\put(690.0,192.0){\rule[-0.200pt]{4.818pt}{0.400pt}}
\put(80.0,279.0){\rule[-0.200pt]{4.818pt}{0.400pt}}
\put(70,279){\makebox(0,0)[r]{1.5}}
\put(690.0,279.0){\rule[-0.200pt]{4.818pt}{0.400pt}}
\put(80.0,367.0){\rule[-0.200pt]{4.818pt}{0.400pt}}
\put(70,367){\makebox(0,0)[r]{2}}
\put(690.0,367.0){\rule[-0.200pt]{4.818pt}{0.400pt}}
\put(80.0,455.0){\rule[-0.200pt]{4.818pt}{0.400pt}}
\put(70,455){\makebox(0,0)[r]{2.5}}
\put(690.0,455.0){\rule[-0.200pt]{4.818pt}{0.400pt}}
\put(80.0,543.0){\rule[-0.200pt]{4.818pt}{0.400pt}}
\put(70,543){\makebox(0,0)[r]{3}}
\put(690.0,543.0){\rule[-0.200pt]{4.818pt}{0.400pt}}
\put(80.0,631.0){\rule[-0.200pt]{4.818pt}{0.400pt}}
\put(70,631){\makebox(0,0)[r]{3.5}}
\put(690.0,631.0){\rule[-0.200pt]{4.818pt}{0.400pt}}
\put(80.0,718.0){\rule[-0.200pt]{4.818pt}{0.400pt}}
\put(70,718){\makebox(0,0)[r]{4}}
\put(690.0,718.0){\rule[-0.200pt]{4.818pt}{0.400pt}}
\put(80.0,806.0){\rule[-0.200pt]{4.818pt}{0.400pt}}
\put(70,806){\makebox(0,0)[r]{4.5}}
\put(690.0,806.0){\rule[-0.200pt]{4.818pt}{0.400pt}}
\put(80.0,894.0){\rule[-0.200pt]{4.818pt}{0.400pt}}
\put(70,894){\makebox(0,0)[r]{5}}
\put(690.0,894.0){\rule[-0.200pt]{4.818pt}{0.400pt}}
\put(80.0,16.0){\rule[-0.200pt]{0.400pt}{4.818pt}}
\put(80,6){\makebox(0,0){\parbox{5ex}{\centering\ \newline 0}}}
\put(80.0,874.0){\rule[-0.200pt]{0.400pt}{4.818pt}}
\put(260.0,16.0){\rule[-0.200pt]{0.400pt}{4.818pt}}
\put(260,6){\makebox(0,0){\parbox{5ex}{\centering\ \newline 0.2}}}
\put(260.0,874.0){\rule[-0.200pt]{0.400pt}{4.818pt}}
\put(440.0,16.0){\rule[-0.200pt]{0.400pt}{4.818pt}}
\put(440,6){\makebox(0,0){\parbox{5ex}{\centering\ \newline 0.4}}}
\put(440.0,874.0){\rule[-0.200pt]{0.400pt}{4.818pt}}
\put(620.0,16.0){\rule[-0.200pt]{0.400pt}{4.818pt}}
\put(620,6){\makebox(0,0){\parbox{5ex}{\centering\ \newline 0.6}}}
\put(620.0,874.0){\rule[-0.200pt]{0.400pt}{4.818pt}}
\put(80.0,16.0){\rule[-0.200pt]{151.767pt}{0.400pt}}
\put(710.0,16.0){\rule[-0.200pt]{0.400pt}{211.510pt}}
\put(80.0,894.0){\rule[-0.200pt]{151.767pt}{0.400pt}}
\put(395,-71){\makebox(0,0){$-\mbhq$}}
\put(-36,455){\makebox(0,0){$\lba$}}
\put(80.0,16.0){\rule[-0.200pt]{0.400pt}{211.510pt}}
\put(80,16){\usebox{\plotpoint}}
\multiput(80.58,16.00)(0.499,0.646){127}{\rule{0.120pt}{0.617pt}}
\multiput(79.17,16.00)(65.000,82.720){2}{\rule{0.400pt}{0.308pt}}
\multiput(145.59,100.00)(0.488,0.560){13}{\rule{0.117pt}{0.550pt}}
\multiput(144.17,100.00)(8.000,7.858){2}{\rule{0.400pt}{0.275pt}}
\multiput(153.59,109.00)(0.489,0.669){15}{\rule{0.118pt}{0.633pt}}
\multiput(152.17,109.00)(9.000,10.685){2}{\rule{0.400pt}{0.317pt}}
\multiput(162.58,121.00)(0.492,0.669){21}{\rule{0.119pt}{0.633pt}}
\multiput(161.17,121.00)(12.000,14.685){2}{\rule{0.400pt}{0.317pt}}
\multiput(174.58,137.00)(0.494,0.625){29}{\rule{0.119pt}{0.600pt}}
\multiput(173.17,137.00)(16.000,18.755){2}{\rule{0.400pt}{0.300pt}}
\multiput(190.58,157.00)(0.496,0.660){41}{\rule{0.120pt}{0.627pt}}
\multiput(189.17,157.00)(22.000,27.698){2}{\rule{0.400pt}{0.314pt}}
\multiput(212.58,186.00)(0.498,0.633){65}{\rule{0.120pt}{0.606pt}}
\multiput(211.17,186.00)(34.000,41.742){2}{\rule{0.400pt}{0.303pt}}
\multiput(246.58,229.00)(0.499,0.652){109}{\rule{0.120pt}{0.621pt}}
\multiput(245.17,229.00)(56.000,71.710){2}{\rule{0.400pt}{0.311pt}}
\multiput(302.58,302.00)(0.497,0.661){53}{\rule{0.120pt}{0.629pt}}
\multiput(301.17,302.00)(28.000,35.695){2}{\rule{0.400pt}{0.314pt}}
\multiput(330.58,339.00)(0.499,0.673){107}{\rule{0.120pt}{0.638pt}}
\multiput(329.17,339.00)(55.000,72.675){2}{\rule{0.400pt}{0.319pt}}
\multiput(385.58,413.00)(0.498,0.667){87}{\rule{0.120pt}{0.633pt}}
\multiput(384.17,413.00)(45.000,58.685){2}{\rule{0.400pt}{0.317pt}}
\multiput(430.58,473.00)(0.498,0.706){75}{\rule{0.120pt}{0.664pt}}
\multiput(429.17,473.00)(39.000,53.622){2}{\rule{0.400pt}{0.332pt}}
\multiput(469.58,528.00)(0.498,0.707){65}{\rule{0.120pt}{0.665pt}}
\multiput(468.17,528.00)(34.000,46.620){2}{\rule{0.400pt}{0.332pt}}
\multiput(503.58,576.00)(0.497,0.729){63}{\rule{0.120pt}{0.682pt}}
\multiput(502.17,576.00)(33.000,46.585){2}{\rule{0.400pt}{0.341pt}}
\multiput(536.58,624.00)(0.497,0.720){61}{\rule{0.120pt}{0.675pt}}
\multiput(535.17,624.00)(32.000,44.599){2}{\rule{0.400pt}{0.338pt}}
\multiput(568.58,670.00)(0.497,0.752){57}{\rule{0.120pt}{0.700pt}}
\multiput(567.17,670.00)(30.000,43.547){2}{\rule{0.400pt}{0.350pt}}
\multiput(598.58,715.00)(0.497,0.735){57}{\rule{0.120pt}{0.687pt}}
\multiput(597.17,715.00)(30.000,42.575){2}{\rule{0.400pt}{0.343pt}}
\multiput(628.58,759.00)(0.497,0.770){53}{\rule{0.120pt}{0.714pt}}
\multiput(627.17,759.00)(28.000,41.517){2}{\rule{0.400pt}{0.357pt}}
\end{picture}
\setlength{\unitlength}{0.240900pt}
\ifx\plotpoint\undefined\newsavebox{\plotpoint}\fi
\begin{picture}(750,900)(0,0)
\font\gnuplot=cmtt10 at 10pt
\gnuplot
\sbox{\plotpoint}{\rule[-0.200pt]{0.400pt}{0.400pt}}%
\put(80.0,16.0){\rule[-0.200pt]{0.400pt}{211.510pt}}
\put(80.0,16.0){\rule[-0.200pt]{4.818pt}{0.400pt}}
\put(70,16){\makebox(0,0)[r]{-3}}
\put(690.0,16.0){\rule[-0.200pt]{4.818pt}{0.400pt}}
\put(80.0,162.0){\rule[-0.200pt]{4.818pt}{0.400pt}}
\put(70,162){\makebox(0,0)[r]{-2.5}}
\put(690.0,162.0){\rule[-0.200pt]{4.818pt}{0.400pt}}
\put(80.0,309.0){\rule[-0.200pt]{4.818pt}{0.400pt}}
\put(70,309){\makebox(0,0)[r]{-2}}
\put(690.0,309.0){\rule[-0.200pt]{4.818pt}{0.400pt}}
\put(80.0,455.0){\rule[-0.200pt]{4.818pt}{0.400pt}}
\put(70,455){\makebox(0,0)[r]{-1.5}}
\put(690.0,455.0){\rule[-0.200pt]{4.818pt}{0.400pt}}
\put(80.0,601.0){\rule[-0.200pt]{4.818pt}{0.400pt}}
\put(70,601){\makebox(0,0)[r]{-1}}
\put(690.0,601.0){\rule[-0.200pt]{4.818pt}{0.400pt}}
\put(80.0,748.0){\rule[-0.200pt]{4.818pt}{0.400pt}}
\put(70,748){\makebox(0,0)[r]{-0.5}}
\put(690.0,748.0){\rule[-0.200pt]{4.818pt}{0.400pt}}
\put(80.0,894.0){\rule[-0.200pt]{4.818pt}{0.400pt}}
\put(70,894){\makebox(0,0)[r]{0}}
\put(690.0,894.0){\rule[-0.200pt]{4.818pt}{0.400pt}}
\put(80.0,16.0){\rule[-0.200pt]{0.400pt}{4.818pt}}
\put(80,6){\makebox(0,0){\parbox{5ex}{\centering\ \newline 0}}}
\put(80.0,874.0){\rule[-0.200pt]{0.400pt}{4.818pt}}
\put(260.0,16.0){\rule[-0.200pt]{0.400pt}{4.818pt}}
\put(260,6){\makebox(0,0){\parbox{5ex}{\centering\ \newline 0.4}}}
\put(260.0,874.0){\rule[-0.200pt]{0.400pt}{4.818pt}}
\put(440.0,16.0){\rule[-0.200pt]{0.400pt}{4.818pt}}
\put(440,6){\makebox(0,0){\parbox{5ex}{\centering\ \newline 0.8}}}
\put(440.0,874.0){\rule[-0.200pt]{0.400pt}{4.818pt}}
\put(620.0,16.0){\rule[-0.200pt]{0.400pt}{4.818pt}}
\put(620,6){\makebox(0,0){\parbox{5ex}{\centering\ \newline 1.2}}}
\put(620.0,874.0){\rule[-0.200pt]{0.400pt}{4.818pt}}
\put(80.0,16.0){\rule[-0.200pt]{151.767pt}{0.400pt}}
\put(710.0,16.0){\rule[-0.200pt]{0.400pt}{211.510pt}}
\put(80.0,894.0){\rule[-0.200pt]{151.767pt}{0.400pt}}
\put(395,-71){\makebox(0,0){$\mrhq$}}
\put(-27,601){\makebox(0,0){$\lre$}}
\put(80.0,16.0){\rule[-0.200pt]{0.400pt}{211.510pt}}
\put(80,894){\usebox{\plotpoint}}
\multiput(80.00,892.94)(2.090,-0.468){5}{\rule{1.600pt}{0.113pt}}
\multiput(80.00,893.17)(11.679,-4.000){2}{\rule{0.800pt}{0.400pt}}
\multiput(95.00,888.93)(1.575,-0.482){9}{\rule{1.300pt}{0.116pt}}
\multiput(95.00,889.17)(15.302,-6.000){2}{\rule{0.650pt}{0.400pt}}
\multiput(113.00,882.92)(1.173,-0.491){17}{\rule{1.020pt}{0.118pt}}
\multiput(113.00,883.17)(20.883,-10.000){2}{\rule{0.510pt}{0.400pt}}
\multiput(136.00,872.93)(0.933,-0.477){7}{\rule{0.820pt}{0.115pt}}
\multiput(136.00,873.17)(7.298,-5.000){2}{\rule{0.410pt}{0.400pt}}
\multiput(145.00,867.93)(0.821,-0.477){7}{\rule{0.740pt}{0.115pt}}
\multiput(145.00,868.17)(6.464,-5.000){2}{\rule{0.370pt}{0.400pt}}
\multiput(153.00,862.93)(0.762,-0.482){9}{\rule{0.700pt}{0.116pt}}
\multiput(153.00,863.17)(7.547,-6.000){2}{\rule{0.350pt}{0.400pt}}
\multiput(162.00,856.92)(0.600,-0.491){17}{\rule{0.580pt}{0.118pt}}
\multiput(162.00,857.17)(10.796,-10.000){2}{\rule{0.290pt}{0.400pt}}
\multiput(174.00,846.92)(0.616,-0.493){23}{\rule{0.592pt}{0.119pt}}
\multiput(174.00,847.17)(14.771,-13.000){2}{\rule{0.296pt}{0.400pt}}
\multiput(190.00,833.92)(0.498,-0.496){41}{\rule{0.500pt}{0.120pt}}
\multiput(190.00,834.17)(20.962,-22.000){2}{\rule{0.250pt}{0.400pt}}
\multiput(212.58,810.83)(0.498,-0.529){65}{\rule{0.120pt}{0.524pt}}
\multiput(211.17,811.91)(34.000,-34.913){2}{\rule{0.400pt}{0.262pt}}
\multiput(246.58,774.54)(0.499,-0.616){109}{\rule{0.120pt}{0.593pt}}
\multiput(245.17,775.77)(56.000,-67.769){2}{\rule{0.400pt}{0.296pt}}
\multiput(302.58,705.27)(0.497,-0.698){53}{\rule{0.120pt}{0.657pt}}
\multiput(301.17,706.64)(28.000,-37.636){2}{\rule{0.400pt}{0.329pt}}
\multiput(330.58,666.06)(0.499,-0.760){109}{\rule{0.120pt}{0.707pt}}
\multiput(329.17,667.53)(56.000,-83.532){2}{\rule{0.400pt}{0.354pt}}
\multiput(386.58,580.85)(0.498,-0.824){87}{\rule{0.120pt}{0.758pt}}
\multiput(385.17,582.43)(45.000,-72.427){2}{\rule{0.400pt}{0.379pt}}
\multiput(431.58,506.72)(0.498,-0.865){77}{\rule{0.120pt}{0.790pt}}
\multiput(430.17,508.36)(40.000,-67.360){2}{\rule{0.400pt}{0.395pt}}
\multiput(471.58,437.45)(0.498,-0.946){67}{\rule{0.120pt}{0.854pt}}
\multiput(470.17,439.23)(35.000,-64.227){2}{\rule{0.400pt}{0.427pt}}
\multiput(506.58,371.41)(0.498,-0.960){65}{\rule{0.120pt}{0.865pt}}
\multiput(505.17,373.21)(34.000,-63.205){2}{\rule{0.400pt}{0.432pt}}
\multiput(540.58,306.21)(0.497,-1.020){61}{\rule{0.120pt}{0.913pt}}
\multiput(539.17,308.11)(32.000,-63.106){2}{\rule{0.400pt}{0.456pt}}
\multiput(572.58,241.26)(0.497,-1.004){61}{\rule{0.120pt}{0.900pt}}
\multiput(571.17,243.13)(32.000,-62.132){2}{\rule{0.400pt}{0.450pt}}
\multiput(604.58,176.99)(0.497,-1.089){57}{\rule{0.120pt}{0.967pt}}
\multiput(603.17,178.99)(30.000,-62.994){2}{\rule{0.400pt}{0.483pt}}
\multiput(634.58,111.92)(0.497,-1.110){55}{\rule{0.120pt}{0.983pt}}
\multiput(633.17,113.96)(29.000,-61.960){2}{\rule{0.400pt}{0.491pt}}
\end{picture}\par\ \\
\caption[O(1): Phasendiagramm]{\label{AB-mbqlba-O1}\fns The phase diagram in
terms of the bare (left) and the renormalized parameters (right).\\
         a) Bare parameters: the system is in the broken phase below the curve
and above the $-\mbhq$--axis. The end of the critical line corresponds to a
maximum value $1/4$ for $x_v$.\\
         b) Renormalized parameters: the system is in the broken phase below
the critical line.\\}
\end{center}
\end{figure}

The underlying $O(1)$--symmetry may be realized in the "symmetric" as well as
in the spontaneously broken phase, characterized by an absolute minimum of the
GEP at $\Phioq>0$. Fig. \ref{AB-mbqlba-O1} (right) shows the phase diagram in
terms of the renormalized parameters. The critical line in coupling constant
space is determined by $\cVG(v)=\cVG(0)$. The range of dimensionless parameters
plotted in Fig. \ref{AB-mbqlba-O1} corresponds to the scaling condition, eq.
(\ref{4Skaling}), which the physical mass has to obey. The system is in the
broken phase for all pairs of coupling constants which lie below the critical
line. There is no symmetric phase for $\mrhq<0$ and no symmetry breaking for
$\lre>0$.

Since renormalization amounts to a finite reparametrization for a given cutoff,
the phase diagram may be represented also in terms of bare parameters. This is
shown in Fig. \ref{AB-mbqlba-O1} (left). The symmetry is spontaneously broken
for all values between the critical line and the $-\mbhq$--axis. This is
similar to the quasiclassical behaviour (tree--level). The relation of the
bare--parameters $\lba$ and $\mbhq$ is almost linear, which results from the
rather linear behaviour of the integrals $\In(x)$ and $\Ie(x)$ within the range
$0<x<\ev$ (cf. Appendix C).\\

\section{Discussion and Conclusions}
Although spontaneous symmetry breaking for the $O(N)$--model has been discussed
before in the framework of the Gaussian Effective Potential \cite{Stevenson87},
only massive excitations were found.
This violation of Goldstone's theorem was ascribed to symmetry
breaking "at the operator level", i.e. current non--conservation
because of the non--zero vacuum expectation value of the field.
As will be demonstrated below, the earlier approach indeed
explicitly breaks the $O(2)$--symmetry. However, the argument concerning
the Noether current needs some refinement. The explicit symmetry
breaking and, equivalently the non--conserved current,
is due to a variational ansatz with two mass parameters. In contrast,
our approach, using one mass parameter, does not have these problems
and we recover Goldstone's theorem. The two different masses emerge,
via a Bogoliubov transformation,
{\em after} shifting the field. This transformation
modifies the trial ground state and, concomitantly, its energy.
Therefore, the ansatz with one mass parameter can  eventually
be superior to the ansatz with two mass parameters.

Let us compare our method to the one of ref.
\cite{Stevenson87} in some more detail. For the Noether
current of the $O(2)$--model one has
\bea
\rho &=& \phi_2 \pi_1 - \phi_1 \pi_2 , \nonumber \\
\vec{j} &=& \phi_1 \nabla \phi_2 - \phi_2 \nabla \phi_1 .
\eea
Since $ \partial_0 \phi_j = \pi_j (j=1,2) $, one generally obtains
\be
\partial_0\rho + \nabla\vec{j} = \phi_2 (\partial_0\pi_1 - \Delta \phi_1)
-\phi_1(\partial_0\pi_2- \Delta \phi_2).
\label{eq:curr}
\ee
It is easy to verify that Heisenberg's equations of motion yield
current conservation. The ansatz in ref. \cite{Stevenson87}
corresponds to the field expansion
\begin{eqnarray}
  \phi_1(\vec{x}) &=& \phi_{0}+Z_{\phi}^{-\frac{1}{2}} \int (dk)_{\Omega_1}
  \lk[ a_{\Omega_1}(\vec{k})e^{i\vec{k} \cdot \vec{x}}+
  a_{\Omega_1}^{\dagger}(\vec{k}) e^{-i\vec{k} \cdot \vec{x}} \rk],\nonumber \\
  \phi_2(\vec{x}) &=& Z_{\phi}^{-\frac{1}{2}} \int (dk)_{\Omega_2}
  \lk[ a_{\Omega_2}(\vec{k})e^{i\vec{k} \cdot \vec{x}}+
  a_{\Omega_2}^{\dagger}(\vec{k}) e^{-i\vec{k} \cdot \vec{x}} \rk],\nonumber \\
  \pi_1(\vec{x})  &=& -iZ_{\phi}^{\frac{1}{2}} \int (dk)_{\Omega_1}
  \omega(\vec{k})
  \lk[ a_{\Omega_1}(\vec{k})e^{i\vec{k} \cdot \vec{x}}-
  a_{\Omega_1}^{\dagger}(\vec{k}) e^{-i\vec{k} \cdot \vec{x}} \rk],\nonumber\\
  \pi_2(\vec{x})  &=& -iZ_{\phi}^{\frac{1}{2}} \int (dk)_{\Omega_2}
  \omega(\vec{k})
  \lk[ a_{\Omega_2}(\vec{k})e^{i\vec{k} \cdot \vec{x}}-
  a_{\Omega_2}^{\dagger}(\vec{k}) e^{-i\vec{k} \cdot \vec{x}} \rk].
\label{eq:Ans22}
\end{eqnarray}
Comparing to our ansatz, cf. eq. (\ref{eq:An2}), we encounter  two
different mass parameters  $\Omega_1$ and $\Omega_2$. It is important
that the variational principle then indeed yields two different values
for the masses  $\Omega_1$ and $\Omega_2$ in case of
a non--zero vacuum expectation value of the scalar field.
Inserting eqs. (\ref{eq:Ans22}) into
eq. (\ref{eq:curr}) gives
\be
\partial_0\rho + \nabla\vec{j} = (\Omega_2^2-\Omega_1^2) \phi_1 \phi_2.
\label{eq:cnc}
\ee
We see that the Noether current associated with the $O(2)$--symmetry
 is  non--conserved for $\phi_{0}\ne 0$.
If the symmetry is unbroken, i.e.  $\phi_{0}=0$, one obtains
$\Omega_1=\Omega_2$ and the current is conserved.
It is evident from eq. (\ref{eq:cnc})
that the ansatz with one mass parameter, eq. (\ref{eq:An2}),
does not violate current conservation, irrespective of the
vacuum expectation value of the field. In this way one can combine
spontaneous symmetry breaking and current conservation, which are both
basic ingredients of the proof of Goldstone's theorem.

It may also be instructive to compare the different approaches
at the level of the effective Hamiltonian which is defined in this work.
We claim that the $O(2)$--symmetry gets broken explicitly
when we use the gap equations of \cite{Stevenson87}.
This can be seen as follows.
Normal ordering of the Hamiltonian with respect to the ground state
defined with two mass parameters \cite{Stevenson87} and a subsequent
subtraction of the vacuum expectation value give
\begin{eqnarray}
 H_{R} &=& H - \int d^{\nu}x V_{G}=  \nonumber \\
       &=& :\int d^{\nu}x   \bigg[ \frac{1}{2}
           Z_{\phi}^{-1}(\pi_{1}^{2}+\pi_{2}^{2})+\frac{1}{2}Z_{\phi}
           ((\nabla\phi_{1})^2+(\nabla\phi_{2})^2)  \nonumber \\
       & & +\frac{1}{2}\lk(m_{B}^{2}+12\lambda_{B}I_{0}^{(1)}
+4\lambda_{B}I_{0}^{(2)}\rk)Z_{\phi} \phi_{1}^{2} \nonumber \\
       & & +\frac{1}{2}\lk(m_{B}^{2}+12\lambda_{B}I_{0}^{(2)}
+4\lambda_{B}I_{0}^{(1)}\rk)Z_{\phi} \phi_{2}^{2}
           +\lambda_{B}Z_{\phi}^{2}\phi^{4}   \bigg]: +C,
\end{eqnarray}
where $C=C(\Omega_1,\Omega_2,\phi_0)$ is an
irrelevant constant and
$I_0^{(j)}=I_0(\Omega_j^2), j=1,2$.
Since only operator identities were inserted and a constant was
subtracted, the Hamiltonian $H_{R}$ is $O(2)$--symmetric. The
transformation properties of the normal ordered operators appearing in $H_{R}$
are remarkable.  For example, an  $O(2)$--rotation with an angle
$\beta$ leads to
\bea
:\phi_{1}^{2}: &\rightarrow& :(\phi_{1}\cos\beta-\phi_{2}\sin\beta)^2 +
\sin^2\beta(I_{0}^{(2)}-I_{0}^{(1)}):, \nonumber \\
:\phi_{2}^{2}: &\rightarrow& :(\phi_{1}\sin\beta-\phi_{2}\cos\beta)^2 +
\sin^2\beta(I_{0}^{(1)}-I_{0}^{(2)}):,
\eea
and consequently the bare mass term $ W_0 := \frac{1}{2}  m_{B}^{2}:\lk[
\phi_{1}^{2}: +\phi_{2}^{2}\rk]: $ is invariant.
Moreover, from the invariance of the original $\phi^4$ term follows
the transformation of the normal ordered quartic term
\bea
:\phi^4: &\rightarrow& :\phi^4 \nonumber \\
&-& (6I_0^{(1)}+2I_0^{(2))}\lk[(\phi_1\cos\beta
-\phi_2\sin\beta)^2-\phi_1^2+\sin^2\beta(I_0^{(2)}-I_0^{(1)})\rk] \nonumber \\
 &-& (2I_0^{(1)}+6I_0^{(2)})\lk[(\phi_1\sin\beta
+\phi_2\cos\beta)^2-\phi_2^2+\sin^2\beta(I_0^{(1)}-I_0^{(2)})\rk]:.
\label{eq:quar}
\eea
Therefore, the term
$ W_1 := \lambda_{B}:\lk[ \phi^{4} + (6I_{0}^{(1)}+2I_{0}^{(2)})\phi_{1}^{2}+
(2I_{0}^{(1)}+6I_{0}^{(2)}) \phi_{2}^{2}\rk] : $
is also invariant. Now consider
the {\em two} gap equations (cf. eq. (\ref{eq:gap2})), which read
\bea
\Omega_{1}^2 &=& m_{B}^{2}+ 4\lambda_{B}
\lk[3I_{0}^{(1)}+I_{0}^{(2)}+3Z_{\phi}\phi_{0}^{2}\rk], \nonumber \\
\Omega_{2}^2 &=& m_{B}^{2}+ 4\lambda_{B}
\lk[I_{0}^{(1)}+3I_{0}^{(2)}+Z_{\phi}\phi_{0}^{2}\rk].
\label{eq:gap22}
\eea
They are supplemented by the $\phi_0$ equation (cf. eq. (\ref{eq:minphi2}))
\begin{equation}
m_{B}^{2}Z_{\phi}\phi_{0}+4\lambda_{B}Z_{\phi}^{2}\phi_{0}^{3}+
4\lambda_{B}Z_{\phi}\phi_{0}(3I_{0}^{(1)}+I_{0}^{(2)})=0.
\label{eq:minphi22}
\end{equation}
These are the variational equations and symmetry properties in the
two mass parameter case.
In ref. \cite{Stevenson87} it is shown that the equations (\ref{eq:gap22},
\ref{eq:minphi22}) indeed support solutions
$\phi_0 \ne 0 , \Omega_1 \ne \Omega_2$ and
for $\phi_0 =0$ one obviously has $\Omega_1 = \Omega_2$.
In the following, the broken phase is considered.
The sum of the invariant terms,
\bea
W=W_0+W_1,
\eea
 appears in the Hamiltonian.
As in section (3), the bare mass and the $I_0$ functions
can be eliminated by means of the variational equations. This simplifies $W$:
\bea
W &=& :\lambda_{B} \phi^{4} +
\lk(\frac{1}{2}\Omega_1^2-6\lambda_{B}\phi_{0}^2\rk)
\phi_{1}^2 + \lk(\frac{1}{2}\Omega_2^2-2\lambda_{B}\phi_{0}^2\rk)\phi_{2}^2 :
\nonumber \\
 &=& :\lambda_{B} \phi^{4} +\frac{1}{2}\Omega_2^2\phi_{2}^2
-2\lambda_{B}\phi_{0}^2\phi^2:.
\label{eq:Wvar}
\eea
Let us now, i.e. after this substitution, check the invariance of $W$.
Obviously, the last term of $W$ is invariant.
A rotation of the quartic term in eq. (\ref{eq:Wvar}), however,
is not cancelled by the corresponding one in the
operator $ \frac{1}{2} \Omega_2^2 \phi_2^2$. In other words, the symmetry
is explicitly broken by the insertion of the variational equations
of ref. \cite{Stevenson87}. On the other hand, one easily verifies that
the procedure followed in our work, i.e. starting with one mass parameter,
preserves the explicit $O(2)$--symmetry (cf. eq. (\ref{eq:Hsy})).
Only after the shift the symmetry is no longer manifest
(see eq. (\ref{eq:Hsny})).  This explains
why we obtain massless bosons, in accordance with Goldstone's theorem.

In this paper, effective Hamiltonians for scalar field theories
have been derived within a variational approach using Gaussian wave
functionals as ansatz for the ground state. The regularized expectation
value of the Hamiltonian has been minimized and, subsequently been subtracted
from the Hamiltonian in order to arrive at an effective theory.
This effective theory can be split into a free Hamiltonian, which already
contains variationally determined parameters, and a residual part.
The latter can be treated perturbatively. In contrast to
perturbation theory from the onset, spontaneous symmetry breaking
and its consequences, are --in principle-- included in such a framework.
Furthermore, the appearance of spontaneous symmetry breaking
is demonstrated beyond the classical level, i.e.
on the level of a self--consistent quantum theory.

The main new results of our work are obtained for the $O(2)$--model,
the standard example for Goldstone's theorem. The appearance of massless
particles is conventionally derived on the classical level. One minimizes
the classical potential and, after shifting the relevant component of
the field, zero mass terms occur in the Lagrangian.
Of course, this does not guarantee massless excitations on the
quantum level. The variational calculation of this paper,
including quantum effects via the Gaussian Effective Potential,
indeed shows spontaneous symmetry breaking.
In the ansatz with one mass parameter the symmetry can only  be
broken by a non--zero expectation value of the field. In this case,
the quadratic part of the effective Hamiltonian is not diagonal
and, in contrast to the one-component case, the mass parameter
is not equal to the physical mass.
Diagonalization of the quadratic Hamiltonian via a unitary Bogoliubov
transformation yields one massive and one massless excitation,
i.e. the Goldstone boson.
Moreover, it has been explicitly shown that taking into
account the one--loop corrections to the effective Hamiltonian, or
equivalently including the remaining terms in lowest order
perturbation theory, the Goldstone boson remains massless.
The analogous calculation in
quantum mechanics (zero space dimension), however, leads to a
finite non--zero frequency.

The expectation value of the Hamiltonian is a priori
ill--defined. Also the equations following from the variational
principle contain infinities. Consequently, the theory has been
regularized.  On a formal level the results are independent of
the chosen scheme. The only assumption is that the variational
equations indeed are satisfied.
Here we have chosen to work with a finite
momentum cutoff and redefined (only) mass and coupling constant.
In contrast to the case where the regulator is
finally removed, spontaneous symmetry breaking occurs. Other
possiblities including a wave function renormalization, e.g.
the autonomous scheme, are not discussed. It should be
emphasized that the formal developments of this
work do not exclude such schemes.

Up to now, the Standard Model is in perfect agreement with
experiment. The only missing link is the Higgs particle, which
has not been detected yet. The simple reason may be that the Higgs
is too heavy for the present energies.
Phenomenologically, a bound on the Higgs--mass is therefore
of crucial importance.
Assuming the value of the symmetry breaking parameter, the condensate,
and interpreting the variational mass of the one--component theory
as the Higgs--mass, we obtain an upper bound for the Higgs--mass of 1.7 TeV.
The two--component scalar theory probably improves this
estimate but we believe that a complete treatment of the Higgs model is
in order. Also from the theoretical side it is interesting to
investigate the Higgs model along the lines given here. A pertinent question
is,
for example, whether it is possible to describe both the Higgs as
well as the Coulomb phase within one gauge. At present,
a study of the Abelian model in the Coulomb gauge is in progress.

\section*{Acknowledgments}
H.W.L.N. was supported in part by the Federal Ministry of Research and
Technology (BMFT) under contract number 06 HD 729. T.G. was supported by
Cusanuswerk. H.W.L.N. acknowledges the hospitality and support
of the Massachusetts Institute of Technology, where part of this work
has been done. Moreover, he thanks A.C. Kalloniatis, A.K. Kerman, F. Lenz, B.
Schreiber and F.M.C. Witte for valuable discussions. \\

\setcounter{equation}{0}
\renewcommand{\theequation}{A.\arabic{equation}}
\section*{Appendix A: Diagonalization of the quadratic Hamiltonian for
arbitrary
angles}
\appendix

In order to diagonalize the Hamiltonian for arbitrary angles it is more
convenient to use the complex field representation
\begin{eqnarray}
  \Phi &=& \sqrt{\frac{1}{2}}(\phi_{1}+i\phi_{2}) ,   \nonumber \\
   \Phi^{\dagger} &=& \sqrt{\frac{1}{2}}(\phi_{1}-i\phi_{2}) ,   \nonumber \\
   \Pi &=& \sqrt{\frac{1}{2}}(\pi_{1}-i\pi_{2}) ,   \nonumber \\
   \Pi^{\dagger}
 &=& \sqrt{\frac{1}{2}}(\pi_{1}+i\pi_{2}) ,
\end{eqnarray}
and the corresponding relations for the creation and annihilation operators.
The quadratic part of the Hamiltonian in terms of these fields reads
\begin{eqnarray}
  H_{0} &=& : \int d^{\nu}x   \bigg[ \Pi^{\dagger}\Pi+
(\nabla\Phi^{\dagger})(\nabla\Phi)
+\Omega^2\Phi^{\dagger}\Phi \nonumber \\
  & &+\frac{1}{2}\Omega^2[e^{2i\alpha}\Phi\Phi+
e^{-2i\alpha}\Phi^{\dagger}\Phi^{\dagger}]  \bigg] :.
\end{eqnarray}
Of course, one can also use the complex representation from the beginning; the
result is the same. In momentum space one has
\begin{eqnarray}
  H_{0} &=&  \int (dk)   \bigg[ \omega(\vec{k})\lk[\tilde{a}^{\dagger}(\vec{k})
\tilde{a}(\vec{k})+\tilde{b}^{\dagger}(\vec{k})\tilde{b}(\vec{k})\rk]+
  \nonumber \\
& & +\frac{\Omega^2}{2\omega(\vec{k})}  \bigg(\frac{1}{2}e^{2i\alpha}\lk[
\tilde{a}^{\dagger}(\vec{k})\tilde{a}^{\dagger}(-\vec{k})+
2\tilde{a}^{\dagger}(\vec{k})\tilde{b}(\vec{k})+
\tilde{b}(\vec{k})\tilde{b}(-\vec{k})\rk] \nonumber \\
& & +\frac{1}{2}e^{-2i\alpha}\lk[
\tilde{a}(\vec{k})\tilde{a}(-\vec{k})+
2\tilde{b}^{\dagger}(\vec{k})\tilde{a}(\vec{k})+
\tilde{b}^{\dagger}(\vec{k})\tilde{b}^{\dagger}(-\vec{k})\rk]  \bigg)   \bigg].
\end{eqnarray}
These creation and annihilation operators are linear combinations of those in
the main text (cf. section (3)).
Anticipating massless particles we first change the normalization
\begin{equation}
\tilde{a}(\vec{k})=\sqrt{2(2\pi)^{\nu}\omega(\vec{k})}a(\vec{k}),
\end{equation}
and analogous expressions for $a^{\dagger},b$ and $b^{\dagger}$.
It should be emphasized that the $ a, a^{\dagger} $ and $b, b^{\dagger}$
are {\em not}
the operators of section (3) (with the same symbol) but depend
linearly on them. (We thought it would be more confusing to introduce
new symbols than dealing with this small inconvenience with respect to
the notation.)
The commutation relations are now independent of $\omega$
\begin{eqnarray}
& [a(\vec{k}),a^{\dagger}(\vec{k'})]=\delta^{\nu}(\vec{k}-\vec{k'}),
& \nonumber \\
& [b(\vec{k}),b^{\dagger}(\vec{k'})]=\delta^{\nu}(\vec{k}-\vec{k'}).&
\label{eq:CAA}
\end{eqnarray}
The measure in the field expansion also needs to be adjusted
\begin{equation}
 \Phi(\vec{x})= \int \{dk\}_{\Omega}
\{ b(\vec{k})e^{i\vec{k} \cdot \vec{x}}+
a^{\dagger}(\vec{k}) e^{-i\vec{k} \cdot \vec{x}} \} ,
\end{equation}
with
\begin{equation}
\{dk\}_{\Omega}=\frac{d^{\nu}k}{\sqrt{2(2\pi)^{\nu}\omega(\vec{k})}}.
\end{equation}
For the Hamiltonian one obtains
\begin{eqnarray}
  H_{0} &=&  \int d^{\nu}k   \bigg[ \omega(\vec{k})\lk[a^{\dagger}(\vec{k})
a(\vec{k})+b^{\dagger}(\vec{k})b(\vec{k})\rk] \nonumber \\
& &  +\frac{\Omega^2}{2\omega(\vec{k})}  \bigg(\frac{1}{2}e^{2i\alpha}\lk[
a^{\dagger}(\vec{k})a^{\dagger}(-\vec{k})+
2a^{\dagger}(\vec{k})b(\vec{k})+
b(\vec{k})b(-\vec{k})\rk] \nonumber \\
& &  +\frac{1}{2}e^{-2i\alpha}\lk[
a(\vec{k})a(-\vec{k})+
2b^{\dagger}(\vec{k})a(\vec{k})+
b^{\dagger}(\vec{k})b^{\dagger}(-\vec{k})\rk]  \bigg)   \bigg].
\label{eq:HA}
\end{eqnarray}
In order to diagonalize we make  the ansatz for the operators $C$ and $D$
\begin{eqnarray}
  C(\kv)       &=& \alpha_1 a(\kv)+\alpha_2 \ad(-\kv)+
                   \beta_1  b(\kv)+\beta_2  \bd(-\kv),\nonumber\\
  C^{\dg}(\kv) &=& \alpha_1^* \ad(\kv)+\alpha_2^* a(-\kv)+
                   \beta_1^*  \bd(\kv)+\beta_2^*  b(-\kv),\nonumber\\
  D(\kv)       &=& \gamma_1 a(\kv)+\gamma_2 \ad(-\kv)+
                   \delta_1 b(\kv)+\delta_2 \bd(-\kv),\nonumber\\
  D^{\dg}(\kv) &=& \gamma_1^* \ad(\kv)+\gamma_2^* a(-\kv)+
                   \delta_1^* \bd(\kv)+\delta_2^* b(-\kv) ,
\label{eq:BOA}
\end{eqnarray}
where the $\alpha_{j},\beta_{j},\gamma_{j},\delta_{j}$ are complex c--number
functions of $k,\Omega$ and $\alpha$. The new operators fulfill the standard
commutation relations
\begin{eqnarray}
  \lk[C(\kv),C^{\dg}(\kv')\rk] &=& \delta^{\nu}(\kv-\kv'),\nonumber\\
  \lk[D(\kv),D^{\dg}(\kv')\rk] &=& \delta^{\nu}(\kv-\kv').
\label{eq:CCA}
\end{eqnarray}
Since the Hamiltonian should be diagonal in these operators, one has
\begin{eqnarray}
  \lk[C(\kv),H_0\rk] &=& \eps_1(\kv)C(\kv),\nonumber\\
  \lk[D(\kv),H_0\rk] &=& \eps_2(\kv)D(\kv),
\end{eqnarray}
where $\epsilon_{1,2}(\vec{k})$ are the energy eigenvalues. Inserting
eq. (\ref{eq:BOA}) yields an algebraic eigenvalue
equation for each $\vec{k}$. Its
solution gives the energies
\begin{eqnarray}
  \eps_1(\kv) &=& (\kvq+2\Omq)^{\eh} = \eps(k), \nonumber\\
  \eps_2(\kv) &=& (\kvq)^{\eh} = k,
\end{eqnarray}
and the complex functions
\begin{eqnarray}
  \alpha_1 =& \ev\sqrt{2}e^{-i\alpha}\frac{\eps+\om}{\sqrt{\eps\om}}\qquad
  \alpha_2 &= \ev\sqrt{2}e^{i\alpha}\frac{\eps-\om}{\sqrt{\eps\om}},\nonumber\\
  \beta_1  =& \ev\sqrt{2}e^{i\alpha}\frac{\eps+\om}{\sqrt{\eps\om}}\qquad
  \beta_2  &=
\ev\sqrt{2}e^{-i\alpha}\frac{\eps-\om}{\sqrt{\eps\om}},\nonumber\\
  \gamma_1 =& \ev\sqrt{2}e^{-i\alpha}\frac{k+\om}{\sqrt{k\om}}\qquad
  \gamma_2 &= -\ev\sqrt{2}e^{i\alpha}\frac{k-\om}{\sqrt{k\om}},\nonumber\\
  \delta_1 =& -\ev\sqrt{2}e^{i\alpha}\frac{k+\om}{\sqrt{k\om}}\qquad
  \delta_2 &= \ev\sqrt{2}e^{-i\alpha}\frac{k-\om}{\sqrt{k\om}}.
\end{eqnarray}
These functions determine the transformation, eq. (\ref{eq:BOA}), which,
as one can explicitly check,
indeed diagonalizes the Hamiltonian, eq. (\ref{eq:HA}),
\begin{equation}
  H_0 = \int d^{\nu}k\lk[\eps(k)C^{\dg}(\kv)C(\kv)+kD^{\dg}(\kv)D(\kv)\rk],
\label{eq:HDA}
\end{equation}
where we omitted an irrelevant constant. Moreover, with eqs. (\ref{eq:CAA})
and (\ref{eq:BOA}), the invariance of the canonical commutation relations,
cf. eq. (\ref{eq:CCA}), follows. Note that the diagonalized Hamiltonian,
eq. (\ref{eq:HDA}), is independent of the
angle $\alpha$. This a posteriori justifies fixing it to a convenient
value as was done in the main text.

Finally, we want to derive the explicit operator form of the unitary
transformation, eq. (\ref{eq:BOA}). In other words, the operator $U$ is to be
determined such that
\begin{eqnarray}
  C(\kv)       &=& Ua(\kv)U^{\dg},\nonumber\\
  C^{\dg}(\kv) &=& U\ad(\kv)U^{\dg},\nonumber\\
  D(\kv)       &=& Ub(\kv)U^{\dg},\nonumber\\
  D^{\dg}(\kv) &=& U\bd(\kv)U^{\dg}.
\end{eqnarray}
After defining the canonical operators $A$ and $B$ as
\begin{eqnarray}
  A(\kv)       &=& \eh\sqrt{2}\lk(e^{-i\alpha}a(\kv)+e^{i\alpha}b(\kv)\rk)
                   , \nonumber\\
  A^{\dg}(\kv) &=& \eh\sqrt{2}\lk(e^{i\alpha}\ad(\kv)+e^{-i\alpha}\bd(\kv)\rk)
                   , \nonumber\\
  B(\kv)       &=& \eh\sqrt{2}\lk(e^{-i\alpha}a(\kv)-e^{i\alpha}b(\kv)\rk)
                   , \nonumber\\
  B^{\dg}(\kv) &=& \eh\sqrt{2}\lk(e^{i\alpha}\ad(\kv)-e^{-i\alpha}\bd(\kv)\rk)
,
\label{eq:TRA}
\end{eqnarray}
the transformation reads
\begin{eqnarray}
  C(\kv)       &=& \eh\frac{\eps+\om}{\sqrt{\eps\om}}A(\kv)+
                   \eh\frac{\eps-\om}{\sqrt{\eps\om}}A^{\dg}(-\kv)=:
                    A(\kv)\cosh\rho+ A^{\dg}(-\kv)\sinh\rho,\nonumber\\
  C^{\dg}(\kv) &=& \eh\frac{\eps+\om}{\sqrt{\eps\om}}A^{\dg}(\kv)+
                   \eh\frac{\eps-\om}{\sqrt{\eps\om}}A(-\kv)=:
                    A^{\dg}(\kv)\cosh\rho+ A(-\kv)\sinh\rho,\nonumber\\
  D(\kv)       &=& \eh\frac{k+\om}{\sqrt{k\om}}B(\kv)-
                   \eh\frac{k-\om}{\sqrt{k\om}}B^{\dg}(-\kv)=:
                    B(\kv)\cosh\sigma+ B^{\dg}(-\kv)\sinh\sigma,\nonumber\\
  D^{\dg}(\kv) &=& \eh\frac{k+\om}{\sqrt{k\om}}B^{\dg}(\kv)-
                   \eh\frac{k-\om}{\sqrt{k\om}}B(-\kv)=:
                    B^{\dg}(\kv)\cosh\sigma+ B(-\kv)\sinh\sigma,
\end{eqnarray}
with
\begin{equation}
  \tanh\rho=\frac{\eps-\om}{\eps+\om},
\end{equation}
and
\begin{equation}
  \tanh\sigma=\frac{\om-k}{\om+k}.
\end{equation}
Consequently
\begin{eqnarray}
   C(\kv)       &=& U_3A(\kv)U_3^{\dg},\nonumber\\
   C^{\dg}(\kv) &=& U_3A^{\dg}(\kv)U_3^{\dg},\nonumber\\
   D(\kv)       &=& U_3B(\kv)U_3^{\dg},\nonumber\\
   D^{\dg}(\kv) &=& U_3B^{\dg}(\kv)U_3^{\dg} ,
\end{eqnarray}
with
\begin{equation}
   U_3 = \exp\lk[-\int d^{\nu}k\lk\{
         \rho(k)\lk(A^{\dg}(\kv)A^{\dg}(-\kv)-A(\kv)A(-\kv)\rk)+
       \sigma(k)\lk(B^{\dg}(\kv)B^{\dg}(-\kv)-B(\kv)B(-\kv)\rk)\rk\}\rk]
                           .  \end{equation}
The transformation given by eq. (\ref{eq:TRA}) is most conveniently
written as the
product of two unitary transformations. With
\begin{eqnarray}
   \tilde{a}(\kv)       :=& e^{-i\alpha}a(\kv)\qquad
   \tilde{b}(\kv)       &:= e^{i\alpha}b(\kv),\nonumber\\
   \tilde{a}^{\dg}(\kv) :=& e^{i\alpha}\ad(\kv)\qquad
   \tilde{b}^{\dg}(\kv) &:= e^{-i\alpha}\bd(\kv) ,
\label{eq:PHA}
\end{eqnarray}
it consists of a rotation over an angle $\vph,\vph=\frac{\pi}{4}$
\begin{eqnarray}
  A(\kv)       &=& \tilde{a}(\kv)\cos\vph+
                   \tilde{b}(\kv)\sin\vph,\nonumber\\
  A^{\dg}(\kv) &=& \tilde{a}^{\dg}(\kv)\cos\vph+
                   \tilde{b}^{\dg}(\kv)\sin\vph,\nonumber\\
  B^{\dg}(\kv) &=& \tilde{a}(\kv)\cos\vph-
                   \tilde{b}(\kv)\sin\vph,\nonumber\\
  B^{\dg}(\kv) &=& \tilde{a}^{\dg}(\kv)\cos\vph-
                   \tilde{b}^{\dg}(\kv)\sin\vph .
\end{eqnarray}
The second unitary transformation therefore is
\begin{equation}
  U_2=\exp\lk[\vph\int d^{\nu}k\lk\{
      \tilde{b}^{\dg}(\kv)\tilde{a}(\kv)-
      \tilde{a}^{\dg}(\kv)\tilde{b}(\kv)\rk\}\rk].
\end{equation}
The first one only yields the phases, cf. eq. (\ref{eq:PHA}),
\begin{equation}
  U_1=\exp\lk[i\alpha\int d^{\nu}k\lk\{
      \tilde{b}^{\dg}(\kv)\tilde{a}(\kv)-
      \tilde{a}^{\dg}(\kv)\tilde{b}(\kv)\rk\}\rk].
\end{equation}
In this way we finally have
\begin{equation}
  U = U_3U_2U_1.
\end{equation}

\setcounter{equation}{0}
\renewcommand{\theequation}{B.\arabic{equation}}
\section*{Appendix B: Infrared Analysis and Comparison}
In order to prove eq. (\ref{eq:EFTF}) and to compare to quantum mechanics a
careful infrared analysis is in order. Of course, the dimensionality
of space is of crucial importance. Consider, for instance, the infrared
behaviour of the integrals $I_{0}(\delta^2)$. For  $\delta \downarrow 0$
one finds
\begin{eqnarray}
I_{0}(\delta^2) &\sim& \delta^{-1} , \; \nu=0, \nonumber \\
I_{0}(\delta^2) &\sim& \log\delta,  \; \nu=1, \nonumber \\
I_{0}(\delta^2) &\sim& \delta , \; \nu=2, \nonumber \\
I_{0}(\delta^2) &\sim& \delta^{2} , \; \nu=3 .
\end{eqnarray}
In quantum mechanics, i.e. for $\nu=0$, we actually have
\begin{equation}
I_{0}(M^2)=\frac{1}{2M}.
\label{eq:I0QM}
\end{equation}

Let us now come back to eq. (\ref{eq:dE2FT}) for $\nu \ge 1$, i.e.
field theory, and introduce the
notation $ A=\sqrt{q^2+M_1^2}$, $ B=\sqrt{q^2+M_2^2}$ .
Rewrite the first integrand as
\begin{equation}
\frac{1}{AB}\left(\frac{1}{M_2-A-B}\right) =
\frac{1}{AB}\left(\frac{-1}{A+B}+\frac{M_2}{(M_2-A-B)(A+B)}\right).
\end{equation}
Herewith, the first contribution can be expressed in terms of the
$I_0$ integrals and a remaining term $R_1$,
\begin{equation}
\frac{16\lambda_{B}^2Z_{\phi}\phi_{\delta}^2}{M_2}
\int\frac{d^{\nu}q}{2(2\pi)^3}
\frac{1}{AB}\frac{1}{M_2-A-B}=\frac{2\lambda_{B}}{M_2}
\left[I_0(M_1^2)-I_0(M_2^2)\right]+ O(\delta\log\delta)+R_1.
\label{eq:BR1}
\end{equation}
The last term explicitly reads
\begin{equation}
R_1=2\lambda_{B}(M_1^2+O(\delta^2)) \int\frac{d^{\nu}q}{2(2\pi)^3}
\frac{1}{AB}\frac{1}{(M_2-A-B)(A+B)}.
\end{equation}
In one space dimension this is logarithmically divergent.
Similar manipulations for the second contribution lead to the
same result as eq. (\ref{eq:BR1}) but with the remainder
\begin{equation}
R_2=2\lambda_{B}(M_1^2+O(\delta^2)) \int\frac{d^{\nu}q}{2(2\pi)^3}
\frac{1}{AB}\frac{1}{(M_2+A+B)(A+B)}.
\end{equation}
Adding $R_1$ and $R_2$ actually improves the infrared behaviour:
\begin{equation}
R_1+R_2=2\lambda_{B}(M_1^2+O(\delta^2)) \int\frac{d^{\nu}q}{2(2\pi)^3}
\frac{1}{AB}\frac{1}{(A+B)}\frac{2M_2}{(M_2^2-(A+B)^2)}.
\label{eq:BR12}
\end{equation}
Thus we obtain an $O(\delta\log\delta)$ contribution for $\nu=1$ and
an $O(\delta)$ contribution for $\nu \ge 2$, respectively.
With eqs. (\ref{eq:BR1}) and (\ref{eq:BR12}), the proof of
eq. (\ref{eq:EFTF}) is now trivially completed.

These infrared arguments cannot be taken over immediately to
quantum mechanics, $\nu=0$. The $O(\delta\log\delta)$ ($\nu=1$)
terms have to be replaced
by $O(1)$ and thus one  cannot exclude finite corrections.
Indeed, these were obtained in the explicit quantum mechanical
calculation. However, one can also derive that result
starting from eq. (\ref{eq:dE2FT}).
Apart from eliminating the basic integral
via eq. (\ref{eq:I0QM}), one only needs evident identifications, e.g.
$M_1 \leftrightarrow \omega_u$.

\setcounter{equation}{0}
\renewcommand{\theequation}{C.\arabic{equation}}
\section*{Appendix C: Integrals $\IN$ at finite cutoff}
\label{App41}
The integrals $\IN$ with finite momentum cutoff read
\bea \IeO &=& \IntnLdk k^2\sqrt{ k^2+\Omq} = \nonumber\\
          &=& \frac{\La^4}{16\piq}\left[\wepxd-\frac{x}{2}\wepx
             -\frac{x^2}{2}\lnfx\right],\nonumber\\
     \InO &=& \IntnLdk\frac{ k^2}{\sqrt{ k^2+\Omq}} = \nonumber\\
          &=& \frac{\Laq}{8\piq}\left[\wepx-x\lnfx\right],\nonumber\\
    \ImiO &=& \IntnLdk\frac{ k^2}{\sqrt{( k^2+\Omq)^3}} = \nonumber\\
          &=& \frac{1}{4\piq}\left[-\frac{1}{\wepx}+\lnfx\right]
\label{4IN-cut}
\eea
\begin{figure}[p]
\begin{center}
\input{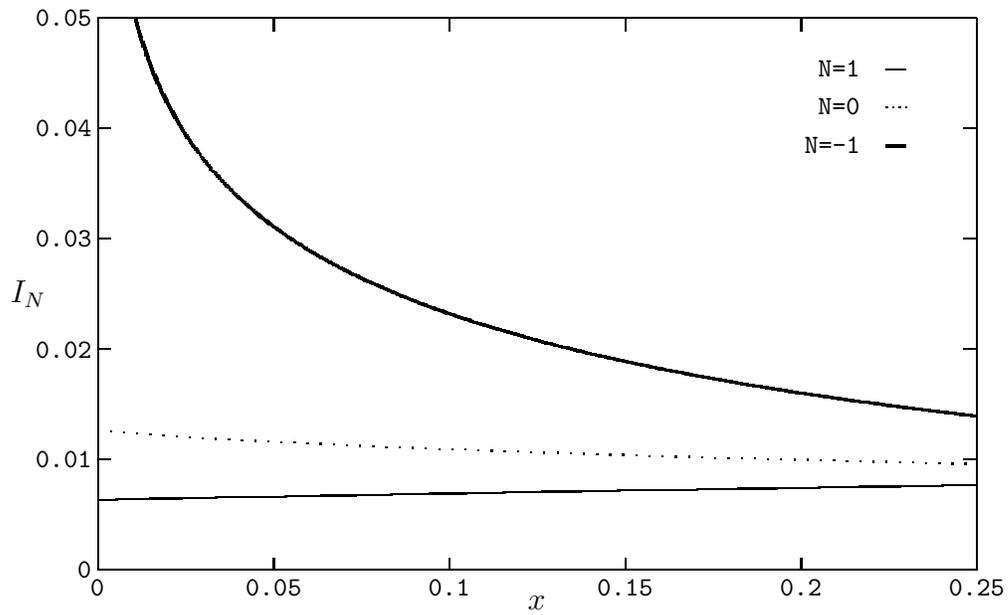}\par\ \\
\caption[The integrals $\IN$]{\label{AB-IN}\fns The integrals $\IN$ as
functions
         of $x=\Omq/\Laq$, for all $x$ which satisfy the scaling condition
         (\ref{4Skaling}).\\}
\end{center}
\end{figure}
($k:=|\kv|$). Fig. \ref{AB-IN} shows the above integrals
($\IN(x):=\IN(\Omq)/\La^{2(N+1)}$) as functions of $x$ for all $x$ which
satisfy the scaling condition (\ref{4Skaling}). The subtraction formula used in
the text,
\be
   \InO-\Inn = -\Omqh\ImiOn+2f'(\Omq), \label{A4Subfor}
\ee
where $\Omn=\Om(\phio=0)$ has been chosen as a reference point, defines the
derivative $\fhxe=d\!\fhx/dx$ of a function $\fh$:
\bea
    \fhxe := \frac{f'(\Omq)}{\Laq}
           &=& \frac{1}{16\piq}   \bigg[\wepx-1-\frac{x}{\wepxn} \nonumber\\
           & & \ \ -x\lnfxxn   \bigg],
\label{4fhxe}
\eea
where $\xn:=\Omnq/\Laq$. With the condition $f(0)=0$ it follows that
\bea
    \fhx := \frac{f(\Omq)}{\La^4}
          &=& \frac{1}{16\piq}   \bigg[\left(1+\frac{x}{2}\right)\wepx-(1+x)
                -\frac{x^2}{2\wepxn} \nonumber\\
          & & \ \ -\frac{x^2}{2}\lnfxxn   \bigg].
\eea
The second derivative will be used as well:
\be
   \fhxz := f''(\Omq) =
   \frac{1}{16\piq}\left[\frac{1}{\wepx}-\frac{1}{\wepxn}-\lnfxxn\right]
\label{4fhxz}
\ee
($\fhxz=d^2\!\fhx/dx^2$, $f''(\Omq)=d^2\!f(\Omq)/d(\Omq)^2$).\\

\setcounter{equation}{0}
\renewcommand{\theequation}{D.\arabic{equation}}
\section*{Appendix D: The case $\Om\equiv 0$}
\label{App42}
The minimization of the GEP by $\Om\equiv 0$ at the nontrivial minimum does not
correspond to a stable minimum of the potential with respect to  $\phio$ {\it
and} $\Om$.
The minimum in this case would be at $\phioq=\vqcl:=-\mrq/4\lre$, which is only
possible for $\mrq<0$. One can easily realize that the extremum is unstable by
considering
\bea
     \left.\dqVGdOq\right|_{\phioq=\vqcl,\Om=0}
 &=& \mrq\Imi(0)<0,\nonumber\\
     \det\,\mbox{\it Hess}\,\left(\VG(\phio,\Om)\right)
 &=& \left|\ba{cc}\dqVGdOq & \dqVGdpdO \\[2mm] \dqVGdOdp & \dqVGdpq \ea
     \right|_{\phioq=\vqcl,\Om=0}\nonumber\\
 &=& -2\mre^4\Imi(0) < 0.
\eea
For $\mrq\ge 0$, the GEP which is minimized by $\Om\equiv 0$ has a minimum only
at $\phio=0$.\\

\end{document}